\documentclass[fleqn,usenatbib]{mnras}

\usepackage{newtxtext,newtxmath}

\usepackage[T1]{fontenc}

\DeclareRobustCommand{\VAN}[3]{#2}
\let\VANthebibliography\thebibliography
\def\thebibliography{\DeclareRobustCommand{\VAN}[3]{##3}\VANthebibliography}

\usepackage{graphicx}	    
\usepackage{amsmath}	    
\usepackage{amssymb}	    
\usepackage{lineno}         

\title[DEM validation for low-gravity simulations]{Validating N-body code \textsc{Chrono} for granular DEM simulations in reduced-gravity environments}

\author[Sunday et al.]{
Cecily Sunday,$^{1,2}$\thanks{E-mail:cecily.sunday@isae-supaero.fr}
Naomi Murdoch,$^{1}$
Simon Tardivel,$^{3}$
Stephen R. Schwartz,$^{4}$
\newauthor \ and Patrick Michel$^{2}$
\\
$^{1}$Department of Electronics, Optics, and Signals, Institut Sup\'erieur de l'A\'eronautique et de l'Espace, 31400 Toulouse, France \\
$^{2}$Universit\'e C\^ote d'Azur, Observatoire de la C\^ote d'Azur, Centre National de la Recherche Scientifique (CNRS), Laboratoire Lagrange, 06300 Nice, France \\
$^{3}$Centre National d'Etudes Spatiales, 31400 Toulouse, France \\
$^{4}$Lunar and Planetary Laboratory, University of Arizona, Tucson, AZ 85721, USA
}

\date{Accepted 2020 August 3. Received 2020 July 4.}

\pubyear{2020}

\begin{document}
\label{firstpage}
\pagerange{\pageref{firstpage}--\pageref{lastpage}}
\maketitle

\begin{abstract}
The Discrete Element Method (DEM) is frequently used to model complex granular systems and to augment the knowledge that we obtain through theory, experimentation, and real-world observations. Numerical simulations are a particularly powerful tool for studying the regolith-covered surfaces of asteroids, comets, and small moons, where reduced-gravity environments produce ill-defined flow behaviors.  In this work, we present a method for validating soft-sphere DEM codes for both terrestrial and small-body granular environments. The open-source code \textsc{Chrono} is modified and evaluated first with a series of simple two-body-collision tests, and then, with a set of piling and tumbler tests. In the piling tests, we vary the coefficient of rolling friction to calibrate the simulations against experiments with 1 mm glass beads.  Then, we use the friction coefficient to model the flow of 1 mm glass beads in a rotating drum, using a drum configuration from a previous experimental study.  We measure the dynamic angle of repose, the flowing layer thickness, and the flowing layer velocity for tests with different particle sizes, contact force models, coefficients of rolling friction, cohesion levels, drum rotation speeds and gravity levels. The tests show that the same flow patterns can be observed at Earth and reduced-gravity levels if the drum rotation speed and the gravity-level are set according to the dimensionless parameter known as the Froude number. \textsc{Chrono} is successfully validated against known flow behaviors at different gravity and cohesion levels, and will be used to study small-body regolith dynamics in future works. 
\end{abstract}




\section{Introduction}
\label{intro}

Past and on-going space missions like NEAR, Dawn, Hayabusa, Rosetta, Hayabusa2, and OSIRIS-REx have provided us with a glimpse into the diverse features found on small-body surfaces \citep{cheng1997, russell2007, fujiwara2006, glassmeier2007, watanabe2019, lauretta2017}. Images show that asteroids are covered with a layer of boulders and regolith, where surface grains vary drastically in terms of size, shape, and material composition \citep{murdoch2015}. Fundamentally, particles interact with one another the same on small bodies as they do on Earth. If an external event agitates a system, grains collide and dissipate energy according to the same contact laws, where their resulting motion depends on collision velocities, internal friction, shape, and material. However, cohesive and electro-static forces are expected to be more influential in reduced-gravity environments than they are on Earth \citep{scheeres2010}, and we are still trying to understand the implications for bulk regolith behavior. 

A limited number of missions have conducted extensive, in-situ operations on small body surfaces. In 2014, the European Space Agency deployed the Philae lander to the surface of comet 67P/Churyumov-Gerasimenko as part of the Rosetta mission \citep{glassmeier2007}. After its landing system failed, Philae rebounded several times on the surface. The lander's bouncing behavior has since been used to characterize the surface mechanical properties of the comet \citep{biele2015}. More recently, the German, French, and Japanese space agencies (DLR, CNES, and JAXA) delivered several hopping-rovers to the surface of the asteroid Ryugu during the Hayabusa2 mission. Data from the rovers and spacecraft are being used to interpret Ryugu's material and geological properties \citep{sugita2019, jaumann2019}. While enlightening, in-situ data is sparse, and additional information is required to explain the phenomena shown in lander and spacecraft images. 

The need to improve our understanding of regolith dynamics in reduced-gravity environments is also important for several up-coming missions. For instance, the Japan Aerospace Exploration Agency's (JAXA) Martian Moons eXploration (MMX) mission \citep{kuramoto2018} will deploy a small rover to the surface of Phobos. The wheeled rover, provided by the Centre National d’Etudes Spatiales (CNES) and the German Aerospace Center (DLR), will operate for three months on Phobos and cover an anticipated distance of a several meters to hundreds of meters \citep{tardivel2019, ulamec2020}. In addition to providing important information regarding the geological and geophysical evolution of Phobos, an understanding of Phobos's surface mechanics is critical to the design and operations of the rover itself. Knowledge of regolith dynamics will also be essential for interpreting the consequences of the impact of NASA's DART mission \citep{cheng2017}, and for preparing the landing of CubeSats on the surface of the asteroid Didymoon during ESA's Hera mission \citep{michel2018}.

Laboratory experiments have been developed to study reduced-gravity impact dynamics \citep{colwell1999, brisset2018, murdoch2017},  avalanching \citep{kleinhans2011, hofmann2017}, angle of repose \citep{nakashima2011}, and dust lofting \citep{hartzell2013, wang2016}. These tests are difficult and costly to run however, as they often rely on parabolic flights, drop-tower set-ups, or shuttle missions to reach variable gravity conditions. As a result, numerical modeling has become an essential tool for studying planetary surfaces. Tests can be carried out across large parameter spaces, and flow behaviors can be analyzed in impressive detail. For example, numerical models have been used to study the strength, re-shaping, and creep stability of rubble-pile asteroids \citep{sanchez2012, sanchez2014, zhang2017, yu2014}. The code \textsc{pkdgrav} \citep{stadel2001, richardson2000} was used to investigate lander-regolith interactions within the context of the Hayabusa2 mission \citep{maurel2018, thuillet2018}, and the code \textsc{ESyS-particle} was used to simulate particle segregation on asteroid surfaces \citep{tancredi2012}.

Codes \textsc{pkdgrav}, \textsc{ESyS-particle}, and the code by S\`anchez and Scheeres simulate granular systems using the Discrete Element Method (DEM). In DEM, surfaces are modeled at the grain level, where various contact laws are used to calculate the kinematics resulting from inter-particle collisions. It is imperative that these models are correctly implemented in the code in order for it to produce reliable results. As such, the goal of this work is to present a robust framework for validating granular DEM codes. The open-source code \textsc{Chrono} is subject to extensive benchmark testing, and is introduced as a strong platform for future use in the planetary science community. In these current code developments, we neglect electrostatic and self-gravity forces, and focus on the effects of rolling resistance, cohesion, and gravity-level. We validate the code by comparing simulations against existing experimental and numerical studies, and by analyzing the flow behavior in a rotating drum in detail. Tumbler experiments have been conducted at increased gravity levels using centrifuges \citep{brucks2007} and reduced gravity levels using parabolic flights \citep{kleinhans2011}. The centrifuge experiments show that the dynamic angle of repose in the drum collapses onto a single curve when plotted against a non-dimensional parameter known as the Froude number. This observation has been reproduced in DEM simulations when $g \geq$ 9.81 m s\textsuperscript{-2} \citep{richardson2011}, but not when $g \leq$ 9.81 m s\textsuperscript{-2}. We show that the same flow patterns exist when $g \leq$ 9.81 m s\textsuperscript{-2}. 

This paper is organized as follows: In Sections \ref{sec:chrono} and \ref{sec:chrono_mods}, we introduce the open-source code used in this study and explain how the code was modified in order to improve its accuracy for reduced-gravity environments. In Section \ref{sec:two_body}, we present a set of low-level tests to evaluate each aspect of the modified code. Then, in Section \ref{sec:pile}, we perform a more complex validation study by comparing experimental and numerical results for a `sand piling' test. Finally, in Section \ref{sec:drum}, we analyze flow behavior in a rotating drum under reduced gravity conditions, with varied particle cohesion. 

\section{Chrono}
\label{sec:chrono}

The simulations presented in this paper are conducted using an open-source dynamics engine called \textsc{Chrono} \citep{tasora2015}. \textsc{Chrono} is used to model either rigid-body or soft-body interactions and can be executed in parallel by using OpenMP, MPI or CUDA algorithms for shared, distributed, or GPU computing \citep{tasora2015, mazhar2013}. Past \textsc{Chrono} studies have spanned a wide range of applications, including structural stability \citep{coisson2016, beatini2017}, 3D printing \citep{mazhar2016}, terrain-vehicle interactions \citep{serban2019}, and asteroid aggregation \citep{ferrari2017}. Given its versatility, users must install the software and construct physical systems based on their individual needs. This study simulates granular systems using the smooth-contact code (SMC) implemented in the \textsc{Chrono::Parallel} module.

\textsc{Chrono::Parallel SMC} follows a traditional soft-sphere DEM (SSDEM) framework, where bodies are considered deformable and are allowed to overlap during collisions. The extent of overlap, relative collision velocity, and other material properties are used to calculate the forces and torques acting on the bodies. Then, particle positions and velocities are updated by resolving all forces and torques in the n-body system (Eqs. \ref{eq:chono_full_trans} and \ref{eq:chono_full_rot}).

\begin{equation}
\label{eq:chono_full_trans}
    m_i \frac{d\vec{v}_i}{dt} = m_i \vec{g} + \sum_{j=1}^{n}\ (\vec{F}_{n} + \vec{F}_{c} + \vec{F}_{t})
\end{equation}

\begin{equation}
\label{eq:chono_full_rot}
    I_i \frac{d\vec{\omega_i}}{dt} = \sum_{j=1}^{n}\ \vec{T}
\end{equation}

\begin{figure}
\begin{centering}
\includegraphics{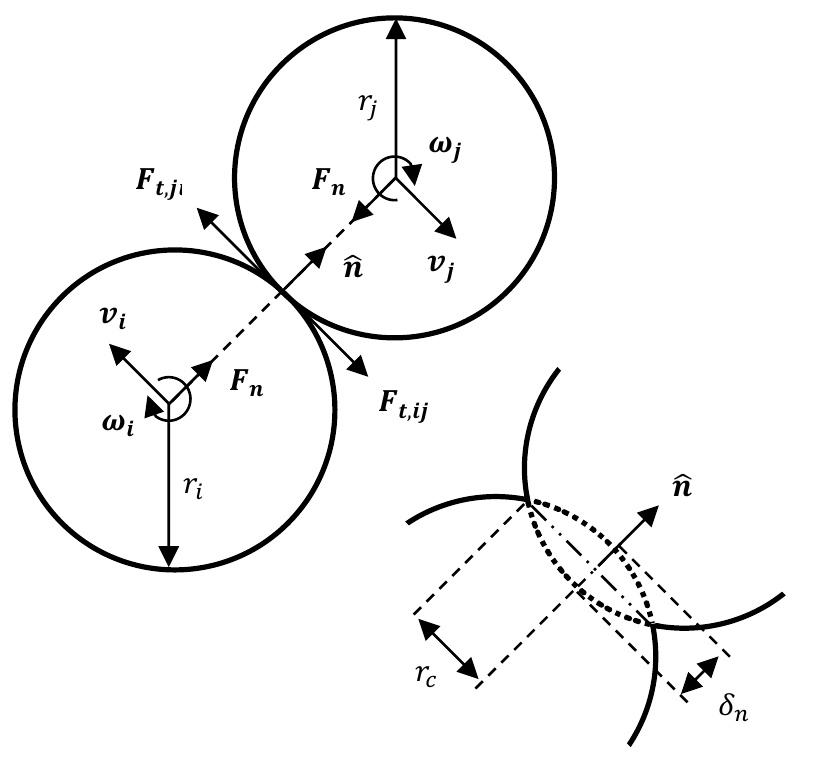}
\caption{Schematic of the interaction between particles `i' and `j'. $\vec{F}_n$ and $\vec{F}_t$ are the normal and tangential forces acting on the bodies, $r$ is the particle radius, $\vec{v}$ is the translational velocity, $\vec{\omega}$ is the rotational velocity, $\delta_n$ is the particle overlap, $r_c$ is the radial distance of the overlapping area, and $\hat{n}$ establishes the contact normal direction. Friction moments and cohesion are not depicted.}
\label{fig:chrono_spheres}
\end{centering}
\end{figure}

In Eqs. \ref{eq:chono_full_trans} and \ref{eq:chono_full_rot}, $m$, $I$, $\vec{v}$ and $\vec{\omega}$ respectively denote particle mass, rotational inertia, translational velocity, and rotational velocity. $\vec{F}_n$, $\vec{F}_c$ and $\vec{F}_t$ are the normal, cohesive and tangential force components, $\vec{g}$ is the acceleration of gravity, and $\vec{T}$ is torque. $\vec{T} = r_i \vec{F}_t \times \hat{n}$, where $r_i$ is particle radius and $\hat{n}$ is the unit vector pointing from one particle center to the other, establishing the contact normal direction (see Fig. \ref{fig:chrono_spheres}). The right hand sides of the equations are summations for all contacts involving particle `i' at the current time, and the equations are applicable to spherical bodies. The specific force and torque models implemented in \textsc{Chrono} are discussed in Sec. \ref{sec:chrono_mods}.

\section{Chrono modifications}
\label{sec:chrono_mods}

Adding rolling and spinning friction to granular DEM simulations has been found to significantly improve a code's ability to replicate bulk granular behaviors like shear strength, dilation response, and shear band development \citep{iwashita1998, mohamed2010}. Since the influence of inter-particle friction becomes even more prominent in airless, reduced-gravity environments, it is essential to consider these components when simulating regolith on small-body surfaces. The SMC code in early versions of \textsc{Chrono::Parallel} (\textsc{Chrono} 4.0.0 and earlier) only considers torques induced by sliding friction and tangential displacement. As part of this work, the SMC code in \textsc{Chrono::Parallel} was updated to incorporate rolling and spinning friction. The code was also modified to include additional force and cohesion models that are relevant for applications in both terrestrial and planetary science. Sections \ref{sec:chrono_normal_force} - \ref{sec:chrono_friction} describe the \textsc{Chrono::Parallel SMC} code in detail, with particular emphasis on the updates available in \textsc{Chrono} 5.0.0.

\subsection{Normal force models}
\label{sec:chrono_normal_force}

In \textsc{Chrono} 4.0.0, normal force $\vec{F}_n$ is calculated using either a Hookean or Hertzian visco-elastic force model, per Eq. \ref{eq:chono_fn}, where $k_n$ is normal stiffness, $g_n$ is normal damping, and $\vec{v}_n$ is the normal component of the relative velocity at the point of contact. Exponents $p$ and $q$ equal 1 and 0 for the Hooke model and 3/2 and 1/4 for the Hertz model \citep{tsuji1992}. $k_n$ and $g_n$ are calculated directly in the \textsc{Chrono} code and depend on user-supplied material properties like Young's modulus, Poisson's ratio, and coefficient of restitution (see Appendix \ref{app:kandg} for more information).

\begin{equation}
\label{eq:chono_fn}
    \vec{F}_{n} = k_n \delta_n^p \hat{n} - g_n \delta_n^q \vec{v}_n
\end{equation}

\begin{equation}
\label{eq:chono_u}
    \vec{v} = \vec{v_j} - \vec{v_i} + r_j (\hat{n} \times \vec{\omega_j}) - r_i (\hat{n} \times \vec{\omega_i})
\end{equation}

Hookean and Hertzian models are commonly implemented in SSDEM codes. However, these models produce a physically unrealistic attractive force during the rebound phase of a collision, when two bodies are about to separate \citep{kruggel2007}. One way to eliminate the non-physical behavior is to set the normal force equal to zero as soon as the force becomes negative, as implemented in \citet{tancredi2012}. The coefficient of restitution associated with the collision can then be estimated using an analytical expression presented in \citet{schwager2008}. The method referenced above is not implemented in \textsc{Chrono}. Instead, the \textsc{Chrono} code includes the option for a third force model that inherently eliminates the non-physical attractive force at the end of the collision. In the model proposed by \citet{flores2011} (Eq. \ref{eq:flores}), the damping component of the normal force calculation is always inferior to the elastic component.  The $c_n$ term is referred to as the hysteresis damping factor and is calculated using Eq. \ref{eq:test_1_flores_gn}, while $e$ is the coefficient of restitution and $v_{o}$ is the initial relative contact velocity between the spheres. 

\begin{equation}
\label{eq:flores}
    \vec{F}_n = k_n \delta^{\,\frac{3}{2}}_n \hat{n} - c_n \delta^{\,\frac{3}{2}}_n \vec{v}_n
\end{equation}
 
 \begin{equation}
    c_n = \frac{8}{5}\ \frac{k_n\ (1 - e)}{e\ v_{o}}
    \label{eq:test_1_flores_gn}
\end{equation}
 
 The advantage of the \citet{flores2011} force model is that it generates a continuous, repulsive force throughout the entirety of the collision. Adding the \citet{flores2011} model to the \textsc{Chrono} 5.0.0 code release allows us to examine how different force models influence simulation results while preserving the user-defined coefficient of restitution value. Sections \ref{sec:two_body_results} and \ref{sec:drum_results_velocity} discuss initial observations on the subject, and a detailed comparison will be carried out as part of future work.

\subsection{Tangential force models}
\label{sec:chrono_tangential_force}
 
Tangential force is limited by the Coulomb friction condition, which establishes a  maximum allowable force $\vec{F}_{t,max}$ given by Eq. \ref{eq:chono_ftmax} \citep{luding2008}.

\begin{equation}
\label{eq:chono_ftmax}
    |\vec{F}_{t,max}| =  \mu_s |\vec{F}_n + \vec{F}_c|
 \end{equation}
 
 The condition relies on the coefficient of static friction $\mu_s$ to define the transition between tangential sticking and slipping. Below the slipping threshold, $\vec{F}_t$ follows Eq. \ref{eq:chono_ftprime}, where $k_t$ is tangential stiffness, $g_t$ is tangential damping, $\vec{v}_t$ is the tangential component of the relative velocity at the point of contact, and $\vec{\delta}_t$ is the tangential displacement vector. Exponents $s$ and $q$ equal 0 for the Hooke model and 1/2 and 1/4 for the Hertz and Flores models, and $k_n$ and $g_n$ are calculated as described in Appendix \ref{app:kandg}. 
 
 \begin{equation}
\label{eq:chono_ftprime}
    \vec{F}_t^\prime = \ -k_t \delta_n^s \vec{\delta}_t - g_t \delta_n^q \vec{v}_{t}
\end{equation}
 
 If $|\vec{F}_t^\prime| \geq |\vec{F}_{t,max}|$, then $|\vec{F}_t| = |\vec{F}_{t,max}|$. Eq. \ref{eq:chono_ft} captures the general form of the tangential force calculation.

\begin{equation}
\label{eq:chono_ft}
    \vec{F}_{t} = min \left[ \mu_s |\vec{F}_n + \vec{F}_c| \frac{\vec{\delta_t}}{ |\delta_t|}, \vec{F}_t^\prime \right]
\end{equation}

The tangential contact displacement vector $\vec{\delta}_t$ is stored and updated at each time step. If $|\vec{F}_t^\prime| \geq |\vec{F}_{t,max}|$, then $\vec{\delta}_t$ is scaled to match the tangential force given by the Coulomb friction condition. \citet{fleischmann2016} describe how $\vec{\delta}_t$ is calculated, updated, and scaled in more detail.

\subsection{Cohesive force models}
\label{sec:chrono_cohesive_force}

\textsc{Chrono} 4.0.0 includes two cohesive force models. The first is a simple model that adds a constant attractive force $C_1$ to all contacting bodies (Eq. \ref{eq:chrono_fc_const}). The second, referred to as the Derjaguin-Muller-Toporov (DMT) model, is dependent on particle effective radius $R$ and an adhesion multiplier $C_2$ (Eqs. \ref{eq:chrono_fc_dmt} and \ref{eq:chrono_r}) \citep{derjaguin1975}. In both models, the cohesive force is applied along the contact normal direction $\hat{n}$.

\begin{equation}
\label{eq:chrono_fc_const}
    \vec{F}_{c,1} = -C_1 \hat{n}
\end{equation}

\begin{equation}
\label{eq:chrono_fc_dmt}
    \vec{F}_{c,2} = -C_2 \sqrt{R}\ \hat{n}
\end{equation}

\begin{equation}
\label{eq:chrono_r}
    R = \frac{r_i r_j}{(r_i + r_j)}
\end{equation}

\textsc{Chrono} 5.0.0 includes the option for a third cohesive force model, given by Eqs. \ref{eq:chrono_fc_perko} and \ref{eq:chrono_fc_perko_c}. The model, based on the work of \citet{perko2001}, was selected for its frequent use in the planetary science field. It accounts for a cleanliness factor $S$, the Hamaker constant $A$, an effective radius $R$, and an inter-particle distance $\Omega$. $S$ is an indicator of the surface separation between two particles at a molecular level. This value approaches unity in low atmospheric pressure or high temperature environments, where the risk of surface contamination from atmospheric gases is greatly reduced \citep{scheeres2010}. The Hamaker constant $A$ is given in units of work (Joules) and is selected according to the material properties of the contacting surfaces. 

\begin{equation}
\label{eq:chrono_fc_perko}
    \vec{F}_{c,3} = -C_3 R\ \hat{n}
\end{equation}

\begin{equation}
\label{eq:chrono_fc_perko_c}
    C_3 = \frac{A S^2}{48\Omega^2}
\end{equation}

\citet{scheeres2010} simplifies Eq. \ref{eq:chrono_fc_perko} for applicability to lunar regolith using $\Omega \sim 1.5 \times 10^{-10}$ m and $A$ = $4.3 \times 10^{-20}$ Joules. Eq. \ref{eq:chrono_fc_scheeres} provides a valid cohesion estimate for Moon-like conditions and a conservative estimate for asteroid and small body surfaces \citep{scheeres2010}. \citet{perko2001} predicts that the cleanliness factor for lunar regolith falls between 0.75 and 0.88.

\begin{equation}
\label{eq:chrono_fc_scheeres}
    \vec{F}_{c,4} = -3.6 \times 10^{-2} S^2 R\ \hat{n}
\end{equation}

\subsection{Friction models}
\label{sec:chrono_friction}

Rolling and spinning frictions are accounted for in the updated code by adjusting the torque calculation to include additional resistance moments. Eq. \ref{eq:chono_full_rot} is replaced by Eq. \ref{eq:chono_full_rotwf},  where $\vec{M}_{r}$ and $\vec{M}_{t}$ are the moments generated by rolling and spinning friction, respectively. The specific resistance models implemented in \textsc{Chrono} 5.0.0 are discussed in Sections \ref{sec:chrono_frict_roll} and \ref{sec:chrono_frict_spin}.

\begin{equation}
\label{eq:chono_full_rotwf}
    I_i \frac{d\vec{\omega_i}}{dt} = \sum_{j=1}^{n}\ (\vec{T} - \vec{M}_{r,i} - \vec{M}_{t,i})
\end{equation}

\subsubsection{Rolling friction}
\label{sec:chrono_frict_roll}

Several rolling resistance models have been explored in past works, including a velocity-independent model \citep{zhou1999}, a viscous model \citep{brilliantov1998, zhou1999}, and various elastic-plastic spring-dashpot models \citep{ai2011, iwashita1998, jiang2005, zhang2017}. The applicability of each model varies based on flow regime and particle shape \citep{ai2011, zhang2017}. Following the approach of \citet{schwartz2012}, \textsc{Chrono} was extended to include a rolling resistance model dependant only on a rolling friction coefficient $\mu_r$, particle radius, normal force magnitude, and the orientation of the rolling axis. The rolling friction torque $\vec{M}_{r,i}$ is calculated per Eq. \ref{eq:chrono_roll} when two particles in persistent contact experience a relative rotational velocity. $\vec{v}_{rot}$ is the relative particle velocity at the point of contact and is calculated according to Eq. \ref{eq:chrono_vrot}.

\begin{equation}
    \label{eq:chrono_roll}
    \vec{M}_{r,i} = {\mu}_{r} r_{i} \frac{(\vec{F}_n \times \vec{v}_{rot})}{|\vec{v}_{rot}|}
\end{equation}

\begin{equation}
    \label{eq:chrono_vrot}
    \vec{v}_{rot} \equiv r_i (\vec{{\omega}}_i \times \hat{n}) - r_j (\vec{{\omega}}_j \times \hat{n}) 
\end{equation}

\subsubsection{Spinning friction}
\label{sec:chrono_frict_spin}

Spinning resistance, also known as twisting resistance, occurs when two bodies in persistent contact rotate at different rates around their contact normal axis $\hat{n}$. As with rolling resistance, spinning resistance can be calculated from either velocity-independent or elastic-plastic spring-dashpot models \citep{schwartz2012, zhang2017}. \textsc{Chrono} 5.0.0 accounts for spinning friction using Eq. \ref{eq:chrono_spin}, where the spinning friction torque $M_{t,i}$, depends on a spinning friction coefficient $\mu_t$, the relative spin velocity between bodies i and j, and the radius of overlap between the two bodies $r_c$ (see Fig. \ref{fig:chrono_spheres} and Eq. \ref{eq:chrono_rc} ) \citep{schwartz2012}.

\begin{equation}
\label{eq:chrono_spin}
    M_{t,i} = {\mu}_{t} r_c \Big [ \frac{(\vec{\omega}_j - \vec{\omega}_i) \cdotp \hat{n}} {|\  \vec{\omega}_j - \vec{\omega}_i\ |} \Big] \vec{F}_n 
\end{equation}

\begin{equation}
\label{eq:chrono_rc}
    r_{c} = \sqrt{r^2_1 - x^2_c}\ \ where\ \ x_{c} = \frac{1}{2} \frac{r^2_i - r^2_j}{(r_i + r_j - {\delta}_n)} + \frac{1}{2} (r_i + r_j - {\delta}_n)
\end{equation}

\section{Two-body validation tests}
\label{sec:two_body}

 DEM codes are often verified against frequently studied problems in granular mechanics, like `sand piling', avalanching, or hopper flow. Specifically, \textsc{Chrono} 4.0.0 was checked against a cone penetration experiment, a direct shear experiment, a standard triaxial test, and a hopper flow experiment \citep{pazouki2017,fleischmann2016}. While these validation efforts returned positive results, the simulations compensated for the code's lack of rolling and spinning resistance by tuning or calibrating the sliding friction parameter to match other experimental results. Large-scale simulations such as these are essential for code validation. However, parameter selection and the bulk behavior of the system can mask low-level issues with the contact models. For this reason, \textsc{Chrono} 5.0.0 was evaluated using seven simple, two-body collision tests before validating the code against more complex systems. 
 
 In general, the two-body tests evaluate interactions between two spheres, a box and a plate, or a sphere and a plate. The tests were influenced by previous validation studies \citep{ai2011, asmar2002, xiang2009, tancredi2012} and were selected to systematically check each aspect of SSDEM implementation in \textsc{Chrono}. When combined, they provide a comprehensive assessment of sliding, rolling, spinning and collision behavior. Sections \ref{sec:two_body_test_1} - \ref{sec:two_body_test_7} describe each test and its associated results in more detail. 
 
\subsection{Simulations and results}
\label{sec:two_body_results}

\begin{figure*}
\includegraphics{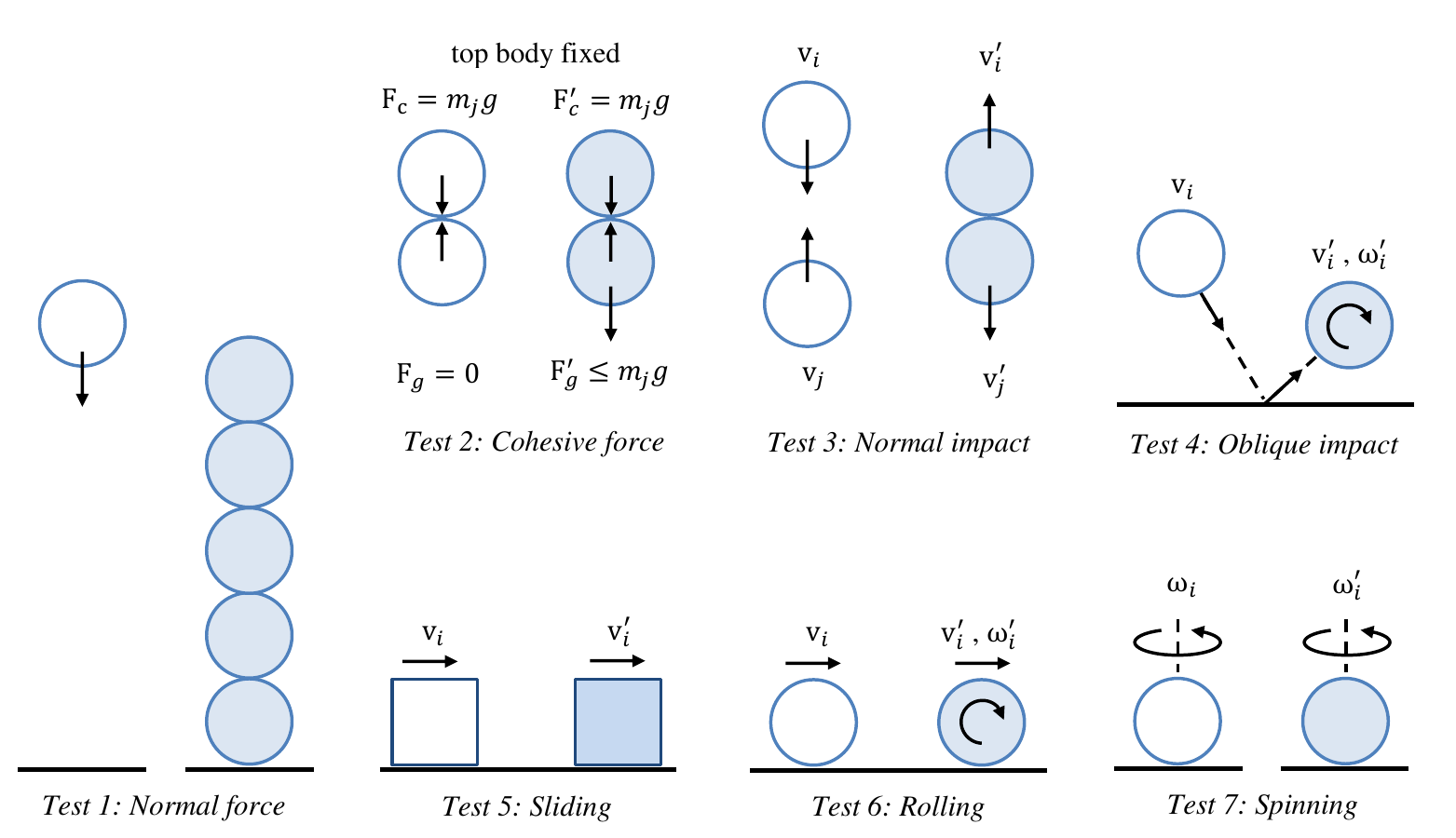}
\caption{Diagram of the seven validation tests presented in Section \ref{sec:two_body}. Bodies on the left depict initial simulation state and bodies on the right illustrate final simulation state. $m$, $g$, $v$, $\omega$, $F_c$, and $F_g$ are respectively particle mass, gravity-level, translational velocity, rotational velocity, cohesive force, and gravitational force.}
\label{fig:two_body} 
\end{figure*}

A visual representation of each test presented in this section can be found in Fig. \ref{fig:two_body}, while all simulation parameters are listed in Table \ref{tab:two_body_params}. If the test involves an interaction between a sphere and a plate or a box and a plate, then the plate is simulated as a large viscoelastic wall with the same material properties as the sphere or box. The simulation time step is determined using the method described later in Section \ref{sec:pile_simsetup} except that the chosen time step is an order of magnitude smaller than required by the calculation. A smaller time step value is applied because the computational cost is negligible. The \textsc{Chrono} 5.0.0 code release passes all seven validation tests. Unless otherwise noted, the results do not vary by force model.

\begin{table*}
\caption{Simulation parameters for the normal force test (Test 1), the cohesive force test (Test 2), the normal impact test (Test 3), the oblique impact test (Test 4), the sliding test (Test 5), the rolling test (Test 6), and the spinning test (Test 7).}
\label{tab:two_body_params}
\begin{tabular}{lcccccccc}
\hline\noalign{\smallskip}
Property & Symbol & Test 1 & Test 2 & Test 3 & Test 4 & Test 5 & Test 6 & Test 7 \\
\noalign{\smallskip}\hline\noalign{\smallskip}
Time step ($\mu s$) & $t$ & 10 & 10 & 10 & 10 & 10 & 10 & 10 \\
Body diameter (m) & $d$ & 1 & 1 & 1 & 1 & 1 & 1 & 1 \\
Body mass (kg) & $m$ & 1 & 1 & 1 & 1 & 1 & 1 & 1 \\
Young's modulus (MPa) & E &  0.5 & 0.5 & 0.5 & 0.5 & 0.5 & 0.5 & 0.5 \\
Poisson's ratio & $\nu$ & 0.3 & 0.3 & 0.3 & 0.3 & 0.3 & 0.3 & 0.3  \\
Static friction coefficient & $\mu_s$ & 0.3 & 0.3 & 0.3 & 0.3 & 0.5 & 0.3 & 0.3 \\
Dynamic friction coefficient & $\mu_k$ & 0.3 & 0.3 & 0.3 & 0.3 & 0.5 & 0.3 & 0.3 \\
Rolling friction coefficient & $\mu_r$ & 0 & 0 & 0 & 0 & 0 & 0.2 & 0 \\
Spinning friction coefficient & $\mu_t$ & 0 & 0 & 0 & 0 & 0 & 0 & 0.2 \\
Cohesive force (N) & $F_c$ & 0 & 10 & 0 & 0 & 0 & 0 & 0 \\
Gravity-level (m s$^{-2}$) & $g$ & 9.81 & 8 - 10 & 0 & 0 & 9.81 & 9.81 & 9.81 \\
Coefficient of restitution & $e$ & 0.3 & 0 & 0 - 1 & 1 & 0 & 0 & 0 \\
\noalign{\smallskip}\hline
\end{tabular}
\end{table*}

\subsubsection{Test 1: Normal force}
\label{sec:two_body_test_1}

The normal force calculation is tested by successively dropping five spheres onto a plate in vertical alignment with one another. Head-on collisions only generate forces in the contact normal direction, so the spheres should not rotate or move laterally when they collide. The test is therefore considered successful if the spheres come to a rest in a stacked position on the plate \citep{asmar2002}. The code passes the test for all three force models.

\subsubsection{Test 2: Cohesive force}
\label{sec:two_body_test_2}

\begin{figure}
\begin{centering}
\includegraphics{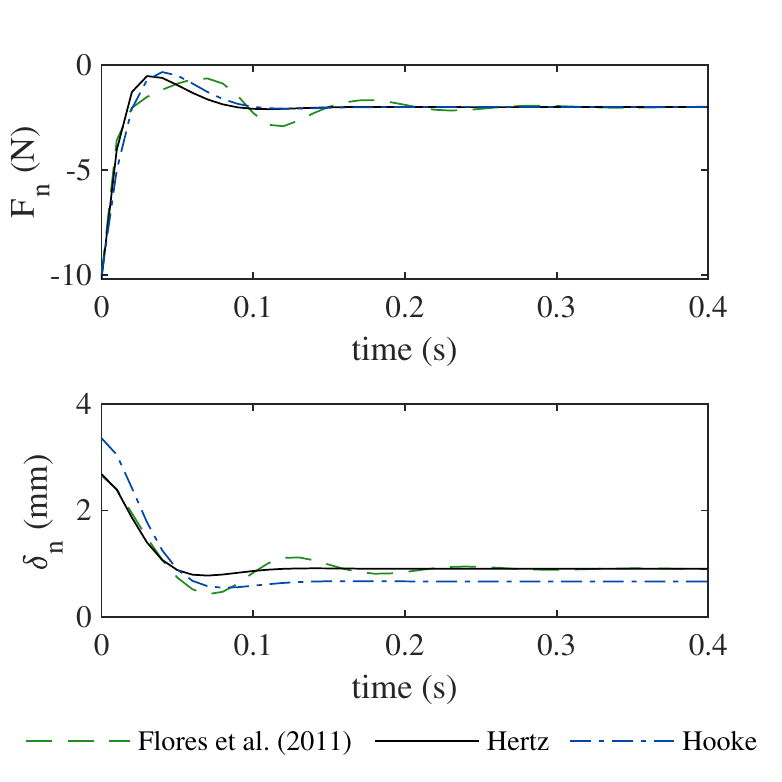}
\caption{Behavior of body `i' during the cohesive force test (Test 2), showing the evolution of normal force $F_n$ (top) and particle overlap $\delta_n$ (bottom) for different force models. When a gravitational force of -8 N is applied to two spheres with a 10 N cohesive force, the bodies stabilize at $F_n = -2$ N without separating.}
\label{fig:two_body_cohesion}
\end{centering}
\end{figure}

The cohesive force calculation is evaluated by applying an external force to two contacting spheres and verifying that the bodies respond in accordance with their cohesive properties. First, two spheres, `i' and `j', are brought into contact, one on top of the other, with `i' on top of `j', in the absence of gravity. Then, sphere `i' is fixed in space, and gravity is applied to the simulation. The gravitational force acts in opposition to the cohesive force between the bodies and attempts to separate sphere `j' from sphere `i'. The test is performed twice; once where the gravitational force is slightly inferior to the cohesive force, and again where the gravitational force equals the cohesive force. The test is considered successful if the spheres remain in contact when $F_c > m_jg$ but separate when $F_c \leq m_jg$. The spheres are expected to separate in the second case due to the elastic nature of the normal force model and the overshoot that occurs when gravity is abruptly turned on. \textsc{Chrono} passes the cohesion test. Overshoot and damping behavior vary by model, as expected (see Fig. \ref{fig:two_body_cohesion}).

\subsubsection{Test 3: Normal impact}
\label{sec:two_body_test_3}

The normal coefficient of restitution $e$ is assessed by observing the rebound behavior of two impacting bodies. Per Eq. \ref{eq:two_body_cor_a}, $e$ is the ratio of the post-collision relative sphere velocity $\vec{v}^\prime$ to the pre-collision relative sphere velocity $\vec{v}$.

\begin{equation}
\label{eq:two_body_cor_a}
    e = \frac{ \vec{v}^\prime}{ \vec{v} }
\end{equation}

In this test, two spheres are positioned and provided with equal and opposite velocities so that they collide head-on. The test is repeated 100 times, with a user-specified $e$ value ranging from zero to one. After each test, the velocities of the spheres are measured, and the output coefficient of restitution $e^\prime$ is calculated according to Eq. \ref{eq:two_body_cor_b}.

\begin{equation}
\label{eq:two_body_cor_b}
    e^\prime = \frac{ \vec{v}^\prime_{i} - \vec{v}^\prime_{j} }{ \vec{v}_{i} - \vec{v}_{j} }
\end{equation}

\begin{figure}
\begin{centering}
\includegraphics{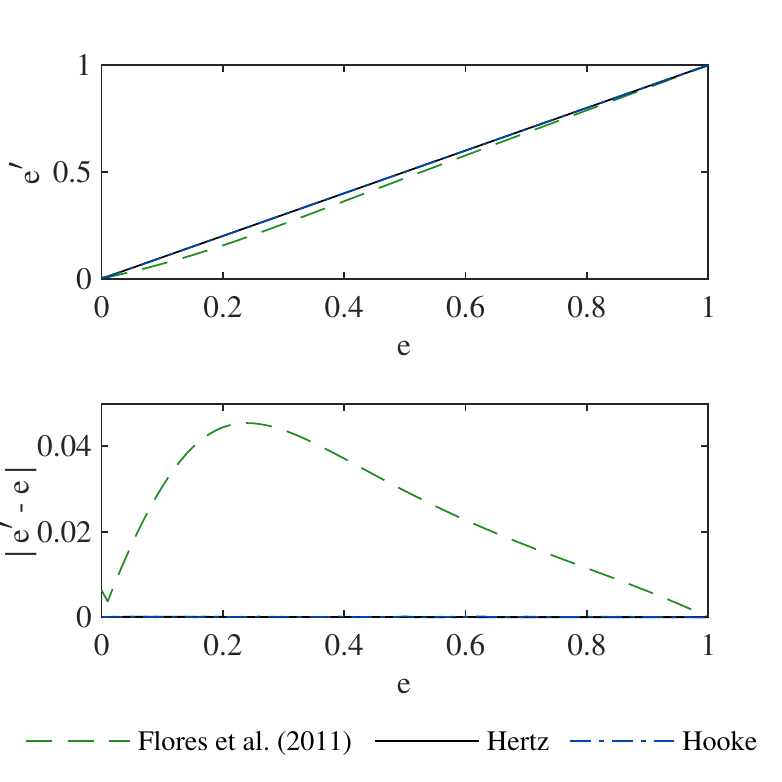}
\caption{Comparison between the expected and measured coefficients of restitution from the normal impact test (Test 3). The differences between $e$ (the expected COR) and $e^\prime$ (the measured COR) are negligible for the Hooke and Hertz models but more important for the \citet{flores2011} model.}
\label{fig:two_body_cor}
\end{centering}
\end{figure}

Based on the implementation of the force models described in Section \ref{sec:chrono_normal_force}, the output coefficient of restitution is expected to match the input value provided by the user. Accordingly, the simulations show that the differences between the expected and measured coefficients of restitution are negligible for the Hooke and Hertz force models. However, the \citet{flores2011} model is only valid for higher coefficient values (see Fig. \ref{fig:two_body_cor}). The differences between $e$ and $e^\prime$ when $e \leq 0.8$ are inherent to the model and do not represent issues with the implementation \citep{flores2011}.

\subsubsection{Test 4: Oblique impact}
\label{sec:two_body_test_4}

The tangential force calculation (Eq. \ref{eq:chono_ft}) is validated by profiling the tangential coefficient of restitution $e_t$ for oblique impacts. A non-rotating sphere is directed toward a plate at impact angles ranging from 2 to 88 degrees. The impact angle $\phi$ is measured from the axis normal to the plate's surface, and gravity is turned off. When the impact angle exceeds a certain threshold, the sphere remains in a sliding regime throughout the duration of the collision. Eq. \ref{eq:two_body_phi} describes the threshold angle $\phi_{th}$ where the collision state transitions from the rolling to sliding regime \citep{yu2017}. 

\begin{equation}
 \label{eq:two_body_phi}
    \phi_{th} = \arctan{\big(\frac{7}{2}\ \mu_s\ (1+e)\big)}
\end{equation}

Once in the sliding regime, $|\vec{F}_t| = \mu_s |\vec{F}_n|$, and certain post-collision properties can be derived using rigid body dynamics \citep{kharaz2001, wu2003, yu2017}. For example, the sphere's post-collision rotational velocity $\omega^\prime_{i}$ and tangential coefficient of restitution $e^\prime_{t}$ can be calculated according to Eqs. \ref{eq:two_body_w} and \ref{eq:two_body_et} respectively, where ${\mu_s}$ is the coefficient of sliding friction, $r_i$ is the radius of the sphere and $\vec{v}_{o,n}$ is the normal component of the initial impact velocity \citep{wu2003}. 

\begin{equation}
\label{eq:two_body_w}
    \omega^\prime_{i} = - \frac{5}{2}\ \frac{\mu_s\ (1+e)\ \vec{v}_{o,n}}{r_i}
\end{equation}

\begin{equation}
\label{eq:two_body_et}
    e^\prime_t = 1 - \frac{\mu_s\ (1+e)}{tan(\phi_{th})}
\end{equation}

The test is considered successful if ${\omega}^\prime_{i}$ and $e^\prime_{t}$ match theoretical results when $\phi \geq \phi_{th}$. For example, when $\mu_s = 0.3$ and $e = 1$, the threshold impact angle for the full sliding regime is 64.54 deg. When $\phi_{th}$ exceeds 64.53 degrees in the simulations, $\omega^\prime_{i}$ = 3 rad s\textsuperscript{-1} and $e^\prime_t$ follows Eq. \ref{eq:two_body_et}. The are no differences between the measured and theoretical values.

\subsubsection{Test 5: Sliding}
\label{sec:two_body_test_5}

In order to test the \textsc{Chrono} implementation of the Coulomb friction condition, a block resting on a plane is provided with an initial horizontal velocity and monitored as it slides across the plane. The block should travel a distance of $d$ given by Eq. \ref{eq:two_body_d} before coming to a rest, where $\mu_s$ is the coefficient of sliding friction, ${v_o}$ is the block's initial horizontal velocity, and $g$ is the acceleration of gravity \citep{xiang2009}.

\begin{equation}
\label{eq:two_body_d}
    d = \frac{v_o^2}{2{\mu_s}g}
\end{equation}

The test is considered successful if the difference between the theoretical and simulated travel distances is less than $1\times 10^{-3}$ m. If ${v_o}$ = 5 m s\textsuperscript{-1} and $\mu_s$ = 0.5, then the block should slide 2.5484 m before coming to a rest. The simulations succeed for these parameters, where the difference between the theoretical and simulated travel distances are $1 \times 10^{-4}$ m, $3 \times 10^{-4}$ m, and $4 \times 10^{-4}$ m for the Hooke, Hertz, and Flores models respectively. 

\subsubsection{Test 6: Rolling}
\label{sec:two_body_test_6}

Rolling resistance is tested by bringing a sphere into contact with a plane, providing it with an initial horizontal velocity, and verifying that it rolls but eventually comes to a rest. The torque generated by rolling friction should be constant and non-zero until the sphere stops rotating. The sphere's position, velocity, and torque profiles should match trends observed in existing works \citep{ai2011, zhou1999}.

In the simulations, the sphere is pushed at an initial velocity of 1 m s\textsuperscript{-1}. Since the sphere is not provided with an initial rotation, it begins by sliding and then starts rolling. Once rolling, the sphere slows down and comes to a rest. The rolling resistance torque is constant while the sphere is in motion.

\subsubsection{Test 7: Spinning}
\label{sec:two_body_test_7}

Spinning resistance is tested by bringing a sphere into contact with a plane, providing the sphere with a rotational velocity around the axis normal to the contact plane, and monitoring the sphere's velocity and torque profiles over time. The test is considered successful if the sphere experiences a constant, non-zero torque while rotating, and if it eventually comes to a rest on the plate.

When the sphere is given an initial spin velocity of 1 rad s\textsuperscript{-1}, it comes to a stop as expected. The spinning resistance torque is constant, but slightly lower for the Hooke model than for the Hertz and Flores models. Since the torque is lower, the sphere takes slightly longer to stop spinning when using the Hooke model. In Eq. \ref{eq:chrono_spin}, we see that spinning resistance is dependant on both normal force and sphere overlap. The spinning torque varies because $\delta_n$ is different for each force model.  

\section{Piling test}
\label{sec:pile}

Piling simulations are ideal for demonstrating the importance of rolling resistance in granular DEM studies. In the past, piling tests have been used to compare different rolling friction models in terms of stability and accuracy \citep{zhou1999, ai2011}, to characterize material properties \citep{zhou2002, li2005}, and to benchmark a code's ability to handle large systems. In this section, we compare experimental and numerical results for a piling test with 1 mm glass beads. The main objective of the test is to ensure that \textsc{Chrono} 5.0.0 functions properly in both flowing and quasi-static states. The pile's angle of repose is used to determine the rolling friction coefficient for the glass beads in the experiment. Then, in Section \ref{sec:pile_results_flow}, the simulated flow is qualitatively compared against theoretical flow behavior in a rectangular hopper.

\subsection{Experimental set-up}
\label{sec:pile_expsetup}

\begin{figure}
\begin{centering}
\includegraphics{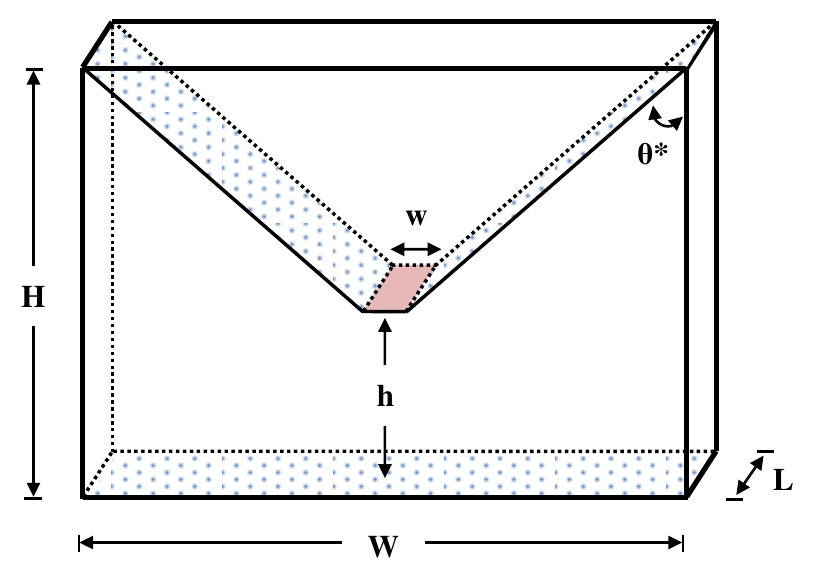}
\caption{Schematic of the experimental container used for the piling experiments and simulations. Patterned surfaces are covered with fixed 1 mm glass beads. The container height $H$ is 177 mm, the container width $W$ is 126 mm, the container length $L$ is 18 mm, the slot width $w$ is 13 mm, the distance from the slot opening to the floor $h$ is 56 mm, and the ramp inclination $\theta^*$ is 50 degrees.}
\label{fig:pile_sketch} 
\end{centering}
\end{figure}

The piling experiment is performed using a thin wooden box with a glass front (see Fig. \ref{fig:pile_sketch}). The box’s internal ramps are angled 50 degrees from vertical to create a 13 mm wide by 18 mm long rectangular slot in the box. The slot remains shut while glass beads are funneled into the box through a hole at the top of the container. Once the particles settle, the slot is manually opened by sliding back a center divider. The beads then flow from the upper portion of the box to the bottom, where they come to rest in a pile. Glass beads are glued to the ramps and floor of the box to increase wall friction, and a Phantom v310 high-speed video camera captures before and after images of the experiment with a 5 mm/pixel spatial resolution. The experiment is repeated six times.

\subsection{Simulation set-up}
\label{sec:pile_simsetup}

The experiment container is re-created in \textsc{Chrono} using plates and spheres. Particles are fixed to the top surfaces of the inclined ramps and floor, mimicking the frictional wall conditions in the experiment. The simulation is executed in two phases: a filling phase and a discharge phase. In the first phase, a funnel is constructed above the container using small, fixed particles. The funnel is filled by arranging particles in a loosely-packed cloud and providing the particles with random initial velocities to promote mixing. The particles fall through the funnel into the container. The filling phase ends when the total kinetic energy of the system falls below $1 \times 10^{-9}$ Joules. This energy level was selected to reduce computation time while ensuring that the simulation ends in a stable state.  In the next phase, the center divider slides back at a rate of 0.1 m s\textsuperscript{-1}, allowing the particles to flow onto the container floor. The simulation ends when the total kinetic energy of the system once again falls below $1 \times 10^{-9}$ Joules.

The parameters used for the piling simulations are listed in Table \ref{tab:pile_dem_params}. Some of the material properties, like density and Poisson’s ratio, map directly to reference sheets for glass beads. The references do not match the exact beads used in the experiment, but are used because the properties should be comparable. 

\begin{table*}
\caption{DEM parameters for simulations replicating glass-bead experiments, such as the piling and tumbler tests presented in Sections \ref{sec:pile} and \ref{sec:drum}.}
\label{tab:pile_dem_params}
\begin{tabular}{llll}
\hline\noalign{\smallskip}
Property & Symbol & Value & Reference  \\
\noalign{\smallskip}\hline\noalign{\smallskip}
Time step ($\mu s$) & $t$ & 1.0 & \citep{huang2014} \\
Particle diameter (mm) & $d$ & 1.0 $\pm$ 0.2 & \\
Particle density (kg m$^{-3}$) & $\rho$ & 2500 & \citep{bolz2019} \\
Young's modulus (MPa) & E & 70 & \citep{bolz2019, chen2017} \\
Poisson's ratio & $\nu$ & 0.24 & \citep{bolz2019} \\
Particle - particle coefficient of restitution & $e_{p}$ & 0.97 & \citep{foerster1994} \\
Particle - wall coefficient of restitution & $e_{w}$ & 0.82 & \citep{alizadeh2014} \\
Particle - particle static friction coefficient & $\mu_{s,p}$ & 0.16 & \citep{alizadeh2014, amstock1997} \\
Particle - wall static friction coefficient & $\mu_{s,w}$ & 0.45  & \citep{alizadeh2014} \\
Particle - particle dynamic friction coefficient & $\mu_{k,p}$ & 0.16  & \citep{alizadeh2014} \\
Particle - wall dynamic friction coefficient & $\mu_{k,w}$ & 0.45 & \citep{alizadeh2014} \\
Rolling friction coefficient & $\mu_{r}$ & 0 - 0.2 &  \\
Spinning friction coefficient & $\mu_{t}$ & 0 &  \\
\noalign{\smallskip}\hline
\end{tabular}
\end{table*}

The simulation time step $t$ is calculated using a conservative estimate for the typical contact duration $t_c$ between two colliding particles in the system. The expression for determining $t_c$ differs by contact model and can depend on parameters like the collision velocity, the material properties of the colliding particles, and even the depth of the particle bed \citep{huang2014}.  some of the simulations described in this study are executed for both the Hookean and Hertzian contact models. A comparison between the Hookean contact time, taken from \citet{schwartz2012}, and the Hertzian contact time, taken from \citet{tancredi2012}, shows that the Hookean model leads to a more conservative (smaller) time step estimate for the simulation configurations described in this paper. Therefore, the contact duration $t_c$ is evaluated for Hookean contact with simple damping, per Eqs. \ref{eq:pile_tc} - \ref{eq:pile_m} \citep{schwartz2012}. Here, $\xi$ is a damping coefficient, $M$ the reduced mass of the contacting particles, $k_n$ is normal stiffness, and $g_n$ is normal damping. The simulation time step is then set as $t = \frac{1}{15}t_c$ to allow for reasonable computation time and sufficient numerical stability.

\begin{equation}
\label{eq:pile_tc}
    t_c = \pi \sqrt{\frac{M}{k_n\ (1 - \xi^2)}}
\end{equation}

\begin{equation}
\label{eq:pile_damping}
    \xi = \frac{g_n}{\sqrt{4M k_n}}
\end{equation}

\begin{equation}
\label{eq:pile_m}
    M = \frac{m_i m_j}{(m_i + m_j)}
\end{equation}

The true value of Young’s modulus, $E$, is estimated to be $\sim$70 GPa for glass beads \citep{bolz2019}. However, large estimates for $E$ result in large stiffness coefficients, short collision durations and the need for unrealistically small time steps (see Appendix \ref{app:kandg} and Eqs. \ref{eq:pile_tc} - \ref{eq:pile_m}). An investigation into the effects of Young’s modulus on particle mixing in a tumbler found that $E$ can be decreased by at least three orders of magnitude before variations in tumbler flow begin to develop \citep{chen2017}. For expediency and consistency between simulations, $E$ was adjusted to 70 MPa for the piling tests. The remaining parameters in Table \ref{tab:pile_dem_params} were selected based on experimental observations from previous works \citep{alizadeh2014, amstock1997, chen2015, foerster1994}.

\subsection{Data processing}
\label{sec:pile_processing}

The experimental angle of repose is estimated using the image processing toolbox in Matlab. First, test images are contrasted and converted into binary format. Then, background noise and pixels belonging to the container are removed. The tail-ends and center of the heap are also identified and removed so that their curvatures do not influence the angle measurement. Finally, the left and right repose angles are determined by fitting lines through the upper edges of the remaining pile.

The simulated angles of repose are found by flattening the final positions of the particles into a 2D plane and fitting a line through the upper-most bodies in the pile. As with the experimental data, the tail-end and center portions of the heap are excluded from the line fit.

\subsection{Results and observations}
\label{sec:pile_results}

\begin{figure}
\begin{centering}
\includegraphics{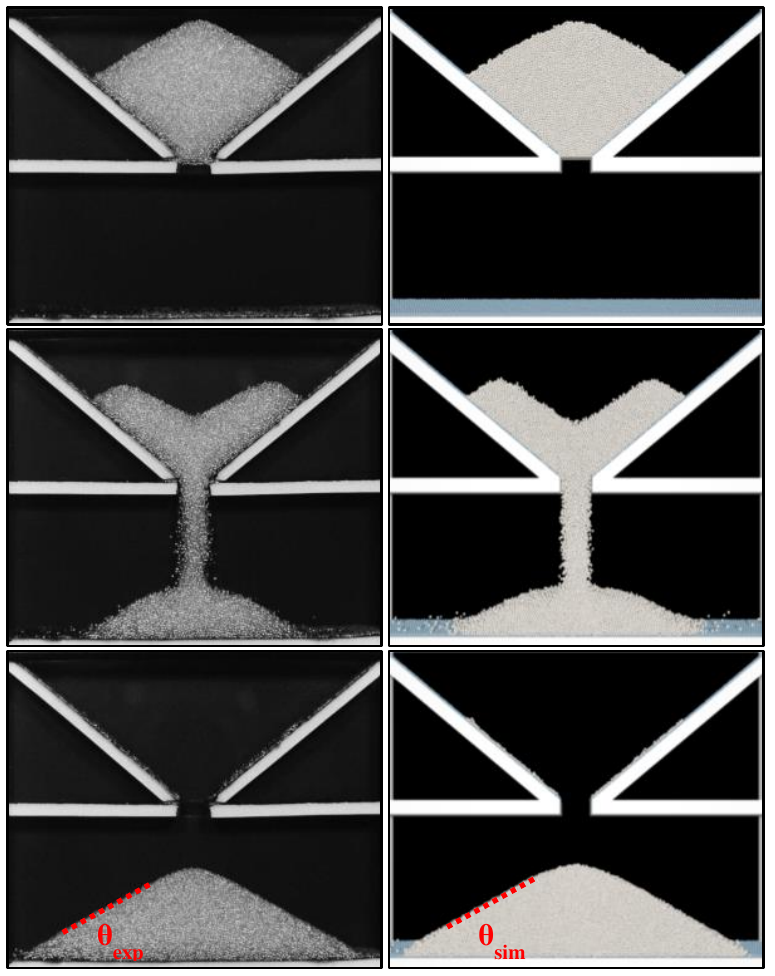}
\caption{Snapshots from the beginning, middle, and end of the piling test, with experiment images on the left and simulation images on the right. The experimental angle of repose $\theta_{exp}$ is 25.2 $\pm$ 0.8 degrees. When $\mu_r$ = 0.09, the simulated angle of repose $\theta_{sim}$ is 25.3 $\pm$ 0.1 degrees.}
\label{fig:pile_comp}
\end{centering}
\end{figure}

Using the method described in Sec. \ref{sec:pile_processing}, the experimental angle of repose was measured as 25.2 $\pm$ 0.8 degrees across six trials. The error represents the standard deviation of the mean from the twelve angle measurements (two measurements, left and right, per trial). Fig. \ref{fig:pile_comp} shows side-by-side snapshots from the real and numerical tests. At a high level, the simulations succeed in reproducing the flow patterns observed in the experiments. Specific details related to angle of repose and flow behavior will be discussed in the Sections \ref{sec:pile_results_theta} and \ref{sec:pile_results_flow}. The simulations contain 58,040 particles and were executed on an Intel\textsuperscript{\textregistered} Xeon\textsuperscript{\textregistered} Gold 6140 processor using 36 OpenMP threads. The discharge phase of the simulations lasted 1.5 real seconds and took approximately 2,000 cpu hours or 2.5 days on a single processor to complete.

\subsection{Angle of repose}
\label{sec:pile_results_theta}

\begin{figure}
\begin{centering}
\includegraphics{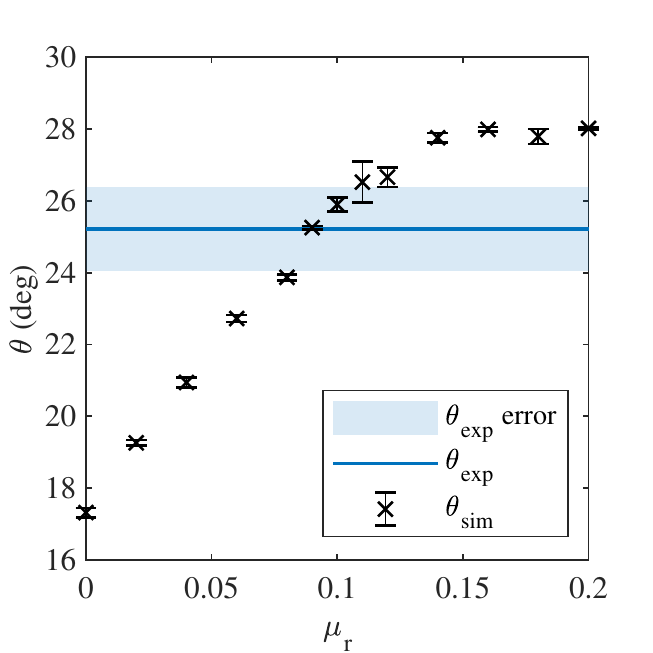}
\caption{Angle of repose $\theta$ for piling simulations with rolling friction coefficient $\mu_r$ ranging from 0 to 0.2. Error bars represent the standard deviation of the mean between the left and right angle measurements. The simulations correspond best with experimental results when $\mu_r$ = 0.09.}
\label{fig:repose} 
\end{centering}
\end{figure}

Glass beads are frequently used for granular testing because their material properties are either well understood or are relatively easy to extract. Certain parameters however, like the coefficients of rolling and spinning friction, are exceptions. Their values are related to specific resistance models, and they are therefore easiest to obtain by calibrating simulations against experimental data. In this study, we vary the coefficient of rolling friction between 0 and 0.2 to find the friction value that most accurately replicates the angle of repose observed in the piling experiments. In Fig. \ref{fig:repose}, we see that the angle of repose increases with rolling friction, and that the simulated pile matches the experimental pile most when $\mu_r = 0.09$. The trend where $\theta$ increases and then levels off is consistent with findings from previous works \citep{zhou1999, zhou2002}.

\subsection{Flow behavior}
\label{sec:pile_results_flow}

The piling test closely resembles the geometry of a rectangular hopper, providing an opportunity to compare simulation data against theoretical flow behaviors. Beverloo et al. (1961) developed a correlation for predicting the mass discharge rate in a cylindrical hopper based on hopper geometry and particle shape \citep{beverloo1961}. Others have since extended the correlation to cover rectangular hoppers \citep{myers1971, brown1965}. Assuming that the hopper width to fill height is sufficiently large, the mass discharge rate $\dot{W}$ is constant and can be calculated using Eq. \ref{eq:pile_discharge}, where $\rho_{flow}$ is the bulk flowing density at the hopper outlet, $k$ is a constant related to particle shape, $w$ is the width of the outlet, $L$ is the length of the outlet, and $\theta^*$ is the hopper angle as measured from vertical \citep{brown1965}.

\begin{equation}
\label{eq:pile_discharge}
    \dot{W} = 1.03\ \rho_{flow}\ g^{1/2}(L - kd)(w - kd)^{3/2}\ (\ \frac{1 - \cos^{3/2}\theta^*}{\sin^{3/2}\theta^*}\ )
\end{equation}

\begin{figure}
\begin{centering}
\includegraphics{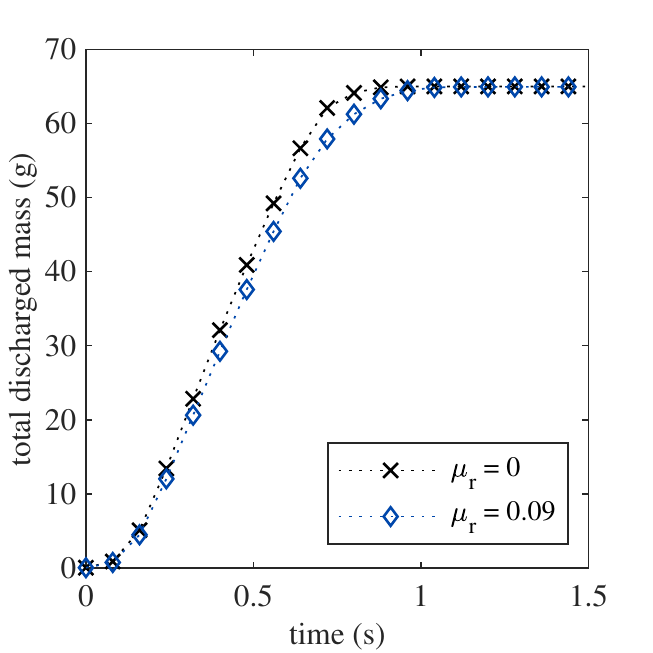}
\caption{Total discharged mass during piling simulations with two different coefficients of rolling friction $\mu_r$.}
\label{fig:pile_discharge} 
\end{centering}
\end{figure}

Unfortunately, it is difficult to calculate the theoretical discharge rate for the piling tests because of the irregularly-shaped surface created by the filling process (see Fig. \ref{fig:pile_comp}). Nonetheless, simulation data can be used to find the total mass discharged over the duration of the simulation. In Fig. \ref{fig:pile_discharge}, we see the rate increase sharply at the beginning of the simulation, remain constant from $t$ = 0.2 s to $t$ = 0.6 s, and then levels off at the end of the simulation. By comparing test cases where $\mu_r = 0$ and $\mu_r = 0.09$, we note that discharge rate decreases slightly as friction increases. 

\begin{figure*}
\includegraphics{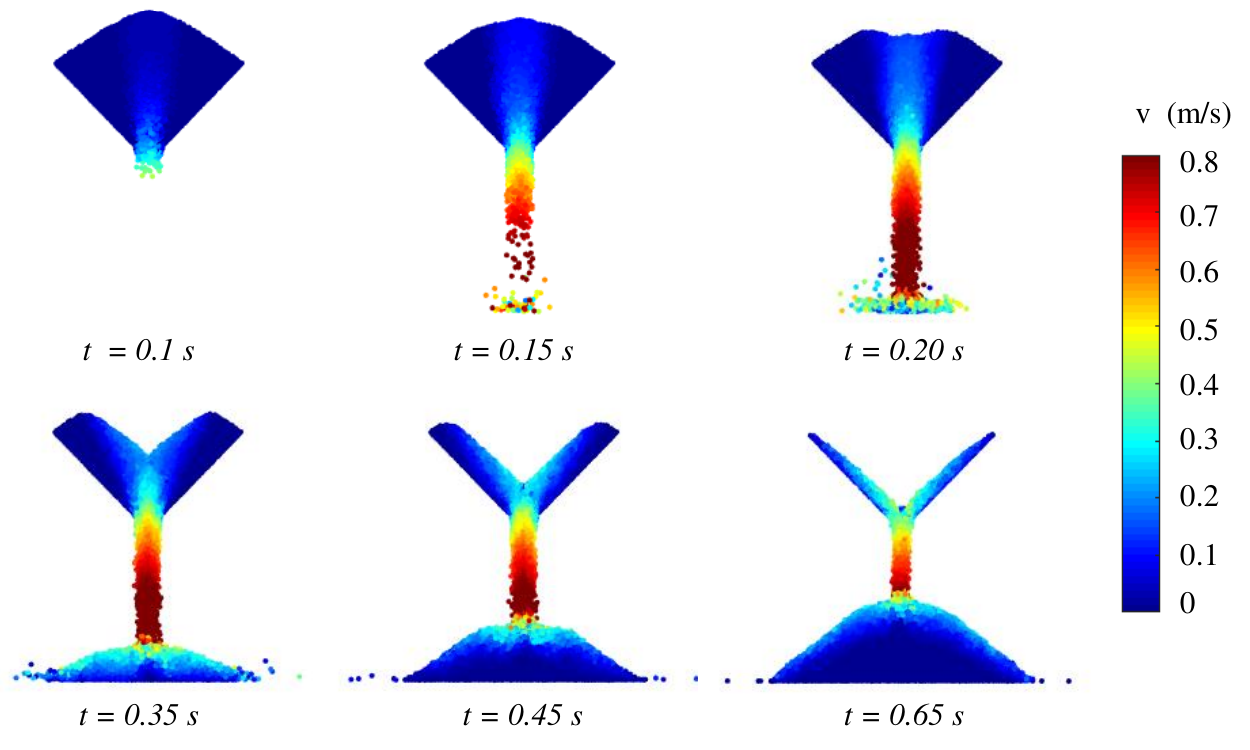}
\caption{Snapshots of particle flow during the piling simulation when $\mu_r = 0.09$. Particles are colored by velocity magnitude.}
\label{fig:pile_vel}
\end{figure*}

Fig. \ref{fig:pile_vel} helps explain the why the mass discharge rate changes as it does. At $t$ = 0.1 s, the slot is only partially open, and the majority of the particle bed are static. Mass discharge increases sharply as the slot opens. From $t$ = 0.15 s to $t$ = 0.35 s, the particles above the slot sink at a uniform speed until they near the orifice. Particle velocities around the orifice increase as the bodies converge and fall through the slot. Mass discharge is nearly constant during this period. At $t$ = 0.45, very few particles remain in the static zone, and the mass discharge rate decreases as the remaining particles exit the system. Qualitatively, the flow matches expected results \citep{anand2008, yan2015, schwartz2012}.

\section{Tumbler flow test}
\label{sec:drum}

Tumbler flow is a significant area of research in granular mechanics. Authors have used rotating drums to study mixing and segregation \citep{gray2005, dury1997, xu2010, chen2017}, to understand the impact of particle size, shape, and friction on flow behavior \citep{santos2016, chou2016, alizadeh2014}, to calibrate material properties in DEM simulations \citep{hu2018}, and to explore the effects of boundary conditions on particle motion \citep{dury1998, felix2002}. Thanks to the abundance of information on the topic, tumbler flow has become a key benchmarking study for DEM code validation. In this section, we numerically replicate tumbler experiments performed by \citet{brucks2007}. We vary the drum rotation speed, gravity-level, particle size, particle friction, and contact model to validate the code against expected behaviors. Then, we use \textsc{Chrono} to investigate the effects of cohesion on flow velocity and regime transitions. 

\subsection{Analytical theory}
\label{sec:drum_theory}

Particles can transition through six flow states in a rotating drum: slipping, slumping, rolling, cascading, cataracting, and centrifuging \citep{henein1983experimental}. The regimes are characterized by different flow patterns, while transitions between the regimes are influenced by parameters like material properties, the tumbler rotation speed, the ratio of drum length to particle diameter (L/d), the ratio of drum diameter to particle diameter (D/d), the drum fill ratio \citep{henein1983modeling, mellmann2001}.  Behavior in the rolling and cascading regimes can be compared in more detail by looking at the flowing layer velocity, flowing layer thickness, and dynamic angle of repose. In this study, the dynamic angle of repose is defined as the angle from horizontal where the surface-layer particles flow at a constant slope (see Fig. \ref{fig:drum_angle_diagram}).

The Froude number, $Fr$, and the granular Bond number, $Bo$, are two dimensionless parameters that are useful for scaling and understating flow behavior in a rotating drum. The Froude number, $Fr$, is the ratio of the centrifugal to the gravitational forces in the tumbler. $Fr$ is calculated according to Eq. \ref{eq:drum_fr}, and depends on drum rotation speed $\omega$, drum diameter $D$, and gravity-level $g$.

\begin{equation}
\label{eq:drum_fr}
    Fr = \frac{\omega^2D}{2g}
\end{equation}

\citet{brucks2007} explore the relationship between gravity-level, drum rotation speed, and angle of repose by conducting a series of tumbler experiments inside of a centrifuge. The authors measure the dynamic angle of repose and the flowing layer thickness for tests with two different drum sizes, drum rotation speeds reaching up to 25 rad s\textsuperscript{-1}, and gravity levels ranging from $1g_o$ to $25g_o$, where $g_o$ is Earth's gravity, or 9.81 m s\textsuperscript{-2}. For more information on the experimental setup, we refer the reader to \citet{brucks2007}. When plotting the angle of repose as a function of Froude number, they found that their data collapses onto a single curve. In the following sections, we perform simulations to see if a similar trend is obtained when $g < 1g_o$ as when $g \geq 1g_o$.

The Froude number is a convenient metric for scaling tumbler flow for different gravity regimes, but a different set of dimensionless numbers is required to describe cohesion-dominated systems. Previous studies looking into the effects of cohesion on tumbler flow have used characterization tools like 1) the collision number, or the ratio of the cohesive to collision forces in the system \citep{nase2001}, 2) the Weber number, or the ratio of the inertial to capillary energy in the system \citep{jarray2017}, 3) the capillary number, or the ratio of the viscous to capillary forces in the system \citep{jarray2017, jarray2019}, and 4) the granular Bond number, or the ratio of the cohesive force to the weight of a single grain in the system. In the following sections, we use the granular Bond number $Bo$ to categorize the level of cohesion in each test configuration (see Eq. \ref{eq:drum_bo}, where $W$ denotes grain weight).

\begin{equation}
\label{eq:drum_bo}
    Bo = \frac{F_c}{W}
\end{equation}

Using the \citet{perko2001} cohesion model, Eq. \ref{eq:drum_bo} becomes Eq. \ref{eq:drum_bo_me}, where $C_3$ is the cohesion multiplier for the \citet{perko2001} cohesion model (Eq. \ref{eq:chrono_fc_perko_c}), $R$ is the effective particle radius, $r$ is the radius of a single particle, and $\rho$ is the density of a single particle.

\begin{equation}
\label{eq:drum_bo_me}
    Bo = \frac{C_3R}{\frac{4}{3}\pi r^3 \rho g}
\end{equation}

Gravity-level and grain size both play key rolls in determining whether a granular system is gravity-dominated or cohesion-dominated. In the absence of moisture content, particles must be sub-millimeter sized or smaller in order for cohesion to influence a granular system on Earth \citep{walton2007}. Conversely, cohesion can in principle become important on small-body surfaces for centimeter sized or larger grains, due to reduced gravity-levels \citep{scheeres2010}. For example, using Eqs. \ref{eq:chrono_fc_scheeres} and \ref{eq:drum_bo} where gravity-level $g$ = 0.0057 m s\textsuperscript{-2} and cleanliness factor $S$ = 0.88, the Bond number for regolith on Phobos, a moon of Mars, nears unity for grains that are approximately 3 cm in diameter. Using different parameters and a notably smaller cleanliness factor, \citet{hartzell2018} finds that cohesive forces come into play for 1 mm or smaller grains on Phobos. In Section \ref{sec:drum_results_cohesion}, we investigate the effects of cohesion on reduced gravity systems by simulated tumbler flow when $g \leq g_o$ and $Bo \geq 1$.

\subsection{Simulation set-up}
\label{sec:drum_setup}

The simulations discussed in this section loosely mimic experiments performed by \citet{brucks2007}, where the drum dimensions match the smaller of the two test set-ups described in the paper. In each simulation, a 60 mm diameter drum is half-filled with particles and rotated at a constant angular velocity. A frictional wall condition is modeled by creating the drum’s inner cylinder out of particles. The inner particle ring rotates as an assembly with the front and back plates. The drum is 5 mm in length and contains either 0.53 $\pm$ 0.05 mm or 1.0 $\pm$ 0.05 mm particles following a normal size distribution. Tests with the smaller particles provide a direct comparison against experimental data, but are computationally expensive due to the large number of particles in the system. Since the \citet{brucks2007} experiments use glass beads, all other simulation parameters are identical to those used in Sec. \ref{sec:pile} (see. Table \ref{tab:pile_dem_params}). 

\begin{figure}
\begin{centering}
\includegraphics{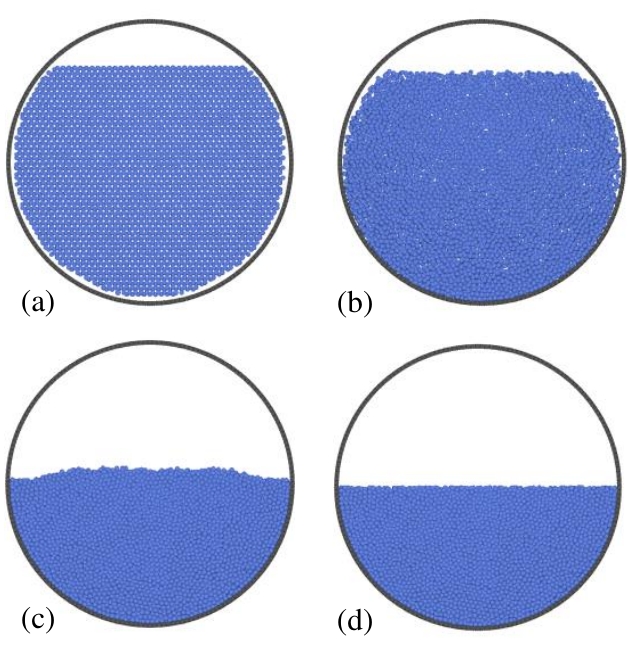}
\caption{Snapshots of the tumbler during simulation set-up. (a) Particles are loosely arranged in the drum, (b) provided with initial velocities so that they mix, and then (c) left to settle under gravity. (d) Particles on the surface of the bed are removed so that the drum is half-filled.}
\label{fig:drum_setup}
\end{centering}
\end{figure}

At the start of the simulation, particles are loosely packed inside of the drum and are provided with random initial velocities in order to generate collisions and promote mixing. Once settled, any particles sitting above the drum’s center line are removed to ensure a half-filled drum-state (see Fig. \ref{fig:drum_setup}). 

Then, the container is rotated at a constant velocity for 5 seconds, or until the system’s total kinetic energy converges to certain value when the particles are in a flowing state. The axis of rotation passes through the center of the container and is parallel to the axis of the cylinder. All simulations were executed using 20-36 OpenMP threads on an Intel\textsuperscript{\textregistered} Xeon\textsuperscript{\textregistered} Gold 6140 processor.

\subsection{Data processing}
\label{sec:drum_processing}

\begin{figure}
\begin{centering}
\includegraphics{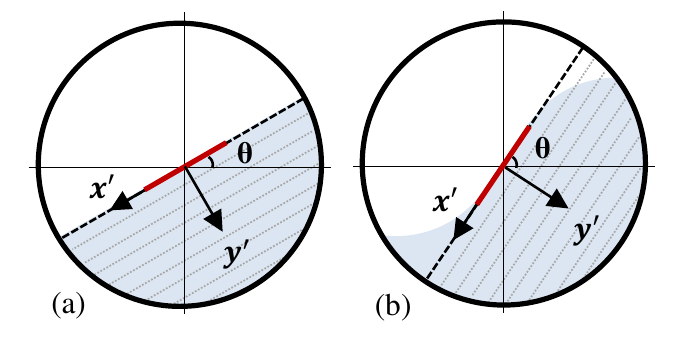}
\caption{Diagram showing the coordinate system used for the analysis of simulations with (a) $Fr = 1\times10^{-4}$ and $Fr = 1\times10^{-3}$ and (b) $Fr = 1\times10^{-2}$ to $Fr$ = 0.1. The angle of repose $\theta$ is measured at the center of the drum (shown by the thick red line), and the streamwise velocities are averaged within the regions running parallel to the $x^\prime$ direction. The spin vector is directed out of page.}
\label{fig:drum_angle_diagram} 
\end{centering}
\end{figure}

Particle positions and velocities are reported in 0.01 second intervals and are used to determine the dynamic angle of repose, the velocity field, and the flowing layer thickness for different test cases. The dynamic angle of repose $\theta$ is calculated from the best-fit line that passes through the top-layer of the particle bed. A mean angle is calculated across 1 simulation second, and the error is reported as the standard deviation of the mean. As Froude number increases, the flowing surface evolves from a flat shape into an S-shaped curve (see Fig. \ref{fig:drum_vel}). The steep angles found at the tail-ends of the S-curve are excluded from the $\theta$ measurement by sampling the position data within a D/2 perimeter about the center of the drum, as shown by the thick red lines in Fig. \ref{fig:drum_angle_diagram}. 

Once the mean repose angle has been measured, particles are binned into regions to construct a streamwise velocity profile. The regions, illustrated in Fig. \ref{fig:drum_angle_diagram} with dotted lines, run parallel to the surface and are approximately two particle diameters thick. Particle velocities in the $x^\prime$ direction are averaged within each region and used to construct a profile for flow velocity $v_x^\prime$ as a function of distance $y^\prime$ from the free surface. Finally, flowing layer thickness $y^\prime_{fl}$ is defined as the distance along $y^\prime$ where the flow reverses direction, indicated by the intersection of the velocity profile with $x^\prime=0$. \citet{alizadeh2014} describe methods for determining $y^\prime_{fl}$ in more detail.

\subsection{Results and observations}
\label{sec:drum_results}

A summary of the tumbler test cases and results are provided in Table \ref{tab:drum_results}. Angle of repose and flowing layer thickness are reported when $Fr$ $\leq$ 0.1. When $Fr > 0.1$, $\theta$ and $y^\prime_{fl}$ cannot be measured because the flow falls into the cataracting and centrifuging regimes. Sections \ref{sec:drum_results_behavior} - \ref{sec:drum_results_velocity} discuss the simulation results in more detail. 

\begin{table*}
\caption{Summary of test cases and results for the tumbler simulations discussed in Section \ref{sec:drum}, where $d$ is the particle diameter, $L$ is the drum length, $\omega$ is the drum rotation speed, $C_3$ is the cohesion multiplier for the \citet{perko2001} cohesion model, $g$ is the gravity-level, $g_o$ is Earth gravity, $Fr$ is the Froude number, $Bo$ is the cohesive bond number, $\mu_r$ is the coefficient of rolling friction, $\theta$ is the dynamic angle of repose, and $y^\prime_{fl}$ is the flowing layer thickness. All other simulation parameters are listed in Table \ref{tab:pile_dem_params}.}
\label{tab:drum_results}
\begin{tabular}{llllllllllllll}
\hline\noalign{\smallskip}
$d$ (mm) & $L/d$ & particles & model & $\omega$ (rpm) & $C_3$ (g s\textsuperscript{-2}) & $g$ & $Fr$ & $Bo$ & $\mu_r$ & $\theta$ (deg) & $y^\prime_{fl}$ \\
\noalign{\smallskip}\hline\noalign{\smallskip}
0.53 & 9.4 & 64 453 & Hertz & 1.7  & 0     & $g_o$       & 0.0001 & 0   & 0.09 & 31.8 $\pm$ 0.8 & 3.5 \\
0.53 & 9.4 & 64 453 & Hertz & 5.4  & 0     & $g_o$       & 0.001  & 0   & 0.09 & 32.7 $\pm$ 0.9 & 4.5 \\
0.53 & 9.4 & 64 453 & Hertz & 17   & 0     & $g_o$       & 0.01   & 0   & 0.09 & 41.1 $\pm$ 0.3 & 6.0 \\
0.53 & 9.4 & 64 453 & Hertz & 39   & 0     & $g_o$       & 0.05   & 0   & 0.09 & 53.5 $\pm$ 0.3 & 7.8 \\
0.53 & 9.4 & 64 453 & Hertz & 55   & 0     & $g_o$       & 0.1    & 0   & 0.09 & 59.2 $\pm$ 0.4 & 9.0 \\
0.53 & 9.4 & 64 453 & Hertz & 122  & 0     & $g_o$       & 0.5    & 0   & 0.09 &                &     \\
0.53 & 9.4 & 64 453 & Hertz & 173  & 0     & $g_o$       & 1.0    & 0   & 0.09 &                &     \\
0.53 & 9.4 & 64 453 & Hertz & 212  & 0     & $g_o$       & 1.5    & 0   & 0.09 &                &     \\
1.0  & 5   & 9 228  & Hertz & 1.7  & 0     & $g_o$       & 0.0001 & 0   & 0.09 & 32.1 $\pm$ 0.6 & 4.2 \\
1.0  & 5   & 9 228  & Hertz & 1.7  & 51.37 & $g_o$       & 0.0001 & 1   & 0.09 & 37.2 $\pm$ 1.7 & 4.6 \\
1.0  & 5   & 9 228  & Hertz & 5.4  & 0     & $g_o$       & 0.001  & 0   & 0    & 26.1 $\pm$ 0.6 & 5.4 \\
1.0  & 5   & 9 228  & Hertz & 5.4  & 0     & $g_o$       & 0.001  & 0   & 0.09 & 35.0 $\pm$ 0.6 & 4.9 \\
1.0  & 5   & 9 228  & Hooke & 5.4  & 0     & $g_o$       & 0.001  & 0   & 0.09 & 31.8 $\pm$ 0.7 & 4.6 \\
1.0  & 5   & 9 228  & Flores& 5.4  & 0     & $g_o$       & 0.001  & 0   & 0.09 & 35.1 $\pm$ 0.7 & 5.0 \\
1.0  & 5   & 9 228  & Hertz & 5.4  & 51.37 & $g_o$       & 0.001  & 1   & 0.09 & 37.1 $\pm$ 0.9 & 4.8 \\
1.0  & 5   & 9 228  & Hertz & 17   & 0     & $g_o$       & 0.01   & 0   & 0.09 & 41.4 $\pm$ 1.1 & 6.5 \\
1.0  & 5   & 9 228  & Hertz & 17   & 51.37 & $g_o$       & 0.01   & 1   & 0.09 & 42.9 $\pm$ 0.7 & 6.2 \\
1.0  & 5   & 9 228  & Hertz & 39   & 0     & $g_o$       & 0.05   & 0   & 0.09 & 51.5 $\pm$ 0.9 & 8.2 \\
1.0  & 5   & 9 228  & Hertz & 39   & 51.37 & $g_o$       & 0.05   & 1   & 0.09 & 53.3 $\pm$ 0.6 & 8.2 \\
1.0  & 5   & 9 228  & Hertz & 55   & 0     & $g_o$       & 0.1    & 0   & 0.09 & 57.2 $\pm$ 0.7 & 9.3 \\
1.0  & 5   & 9 228  & Hertz & 55   & 51.37 & $g_o$       & 0.1    & 1   & 0.09 & 58.3 $\pm$ 0.7 & 9.3 \\
1.0  & 5   & 9 228  & Hertz & 122  & 0     & $g_o$       & 0.5    & 0   & 0.09 &                &     \\
1.0  & 5   & 9 228  & Hertz & 173  & 0     & $g_o$       & 1.0    & 0   & 0.09 &                &     \\
1.0  & 5   & 9 228  & Hertz & 212  & 0     & $g_o$       & 1.5    & 0   & 0.09 &                &     \\
1.0  & 5   & 9 228  & Hertz & 3.8  & 0     & $5g_o$      & 0.0001 & 0   & 0.09 & 32.2 $\pm$ 0.7 & 4.1 \\
1.0  & 5   & 9 228  & Hertz & 3.8  & 256.8 & $5g_o$      & 0.0001 & 1   & 0.09 & 37.3 $\pm$ 2.1 & 4.6 \\
1.0  & 5   & 9 228  & Hertz & 3.8  & 0     & $0.5g_o$    & 0.001  & 0   & 0.09 & 34.9 $\pm$ 0.5 & 5.0 \\
1.0  & 5   & 9 228  & Hertz & 3.8  & 12.84 & $0.5g_o$    & 0.001  & 0.5 & 0.09 & 36.0 $\pm$ 0.6 & 5.0 \\
1.0  & 5   & 9 228  & Hertz & 3.8  & 25.68 & $0.5g_o$    & 0.001  & 1   & 0.09 & 37.0 $\pm$ 0.7 & 4.7 \\
1.0  & 5   & 9 228  & Hertz & 3.8  & 51.37 & $0.5g_o$    & 0.001  & 2   & 0.09 & 39.1 $\pm$ 1.0 & 4.6 \\
1.0  & 5   & 9 228  & Hertz & 3.8  & 77.05 & $0.5g_o$    & 0.001  & 3   & 0.09 & 44.5 $\pm$ 3.1 & 5.1 \\
1.0  & 5   & 9 228  & Hertz & 3.8  & 102.7 & $0.5g_o$    & 0.001  & 4   & 0.09 & 35.0 $\pm$ 2.5 & 4.8 \\
1.0  & 5   & 9 228  & Hertz & 3.8  & 205.5 & $0.5g_o$    & 0.001  & 8   & 0.09 &                &     \\
1.0  & 5   & 9 228  & Hertz & 3.8  & 0     & $0.05g_o$   & 0.01   & 0   & 0.09 & 42.1 $\pm$ 1.6 & 6.5 \\
1.0  & 5   & 9 228  & Hertz & 3.8  & 2.568 & $0.05g_o$   & 0.01   & 1   & 0.09 & 43.1 $\pm$ 0.4 & 6.3 \\
1.0  & 5   & 9 228  & Hertz & 3.8  & 0     & $0.01g_o$   & 0.05   & 0   & 0.09 & 52.3 $\pm$ 0.6 & 8.6 \\
1.0  & 5   & 9 228  & Hertz & 3.8  & 0.503 & $0.01g_o$   & 0.05   & 1   & 0.09 & 53.4 $\pm$ 0.6 & 8.5 \\
1.0  & 5   & 9 228  & Hertz & 3.8  & 0     & $0.005g_o$  & 0.1    & 0   & 0.09 & 57.5 $\pm$ 1.8 & 10.4\\
1.0  & 5   & 9 228  & Hertz & 3.8  & 0.252 & $0.005g_o$  & 0.1    & 1   & 0.09 & 58.6 $\pm$ 1.5 & 10.6\\
1.0  & 5   & 9 228  & Hertz & 3.8  & 0     & $0.001g_o$  & 0.5    & 0   & 0.09 &                &     \\
1.0  & 5   & 9 228  & Hertz & 3.8  & 0     & $0.0005g_o$ & 1.0    & 0   & 0.09 &                &     \\
1.0  & 5   & 9 228  & Hertz & 3.8  & 0     & $0.0003g_o$ & 1.5    & 0   & 0.09 &                &     \\
\noalign{\smallskip}\hline
\end{tabular}
\end{table*}

\subsubsection{Flow behavior}
\label{sec:drum_results_behavior}

\begin{figure*}
\includegraphics{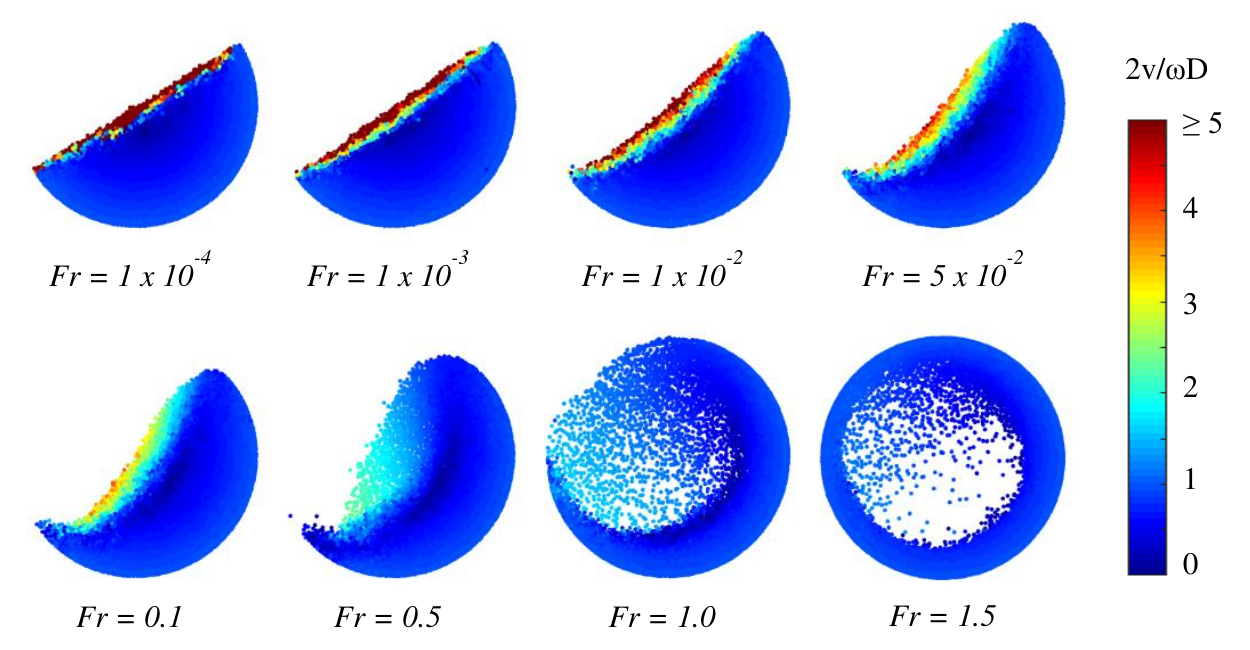}
\caption{Stead-state flow patterns at different Froude numbers for tumbler simulations where $d$ = 1 mm, $\mu_r = 0.09$, and $g = 1g_o$. Particles are colored by velocity magnitude, normalized by drum diameter D and drum rotation speed $\omega$. The flow is in the rolling regime when $Fr = 1\times10^{-4}$ and $1\times10^{-3}$, in the cascading regime when $Fr$ ranges from $1\times10^{-2}$ to 0.1, and in the cataracting regime when $Fr$ = 0.5. The flow is transitioning into the centrifuging regime when $Fr$ = 1.0 and 1.5.}
\label{fig:drum_vel}  
\end{figure*}

Fig. \ref{fig:drum_vel} depicts the evolution of flow behavior with increasing Froude number. Each image represents a snapshot taken at the end of a simulation, where particles are colored by normalized velocity magnitude. At $Fr = 1\times10^{-4}$ and $Fr = 1\times10^{-3}$, the drum motion produces a thin flowing layer at a relatively constant repose angle. The particles in the flowing layer are moving faster than the drum itself, indicating that the flow is in the rolling regime. The flow transitions from the rolling to cascading regime at $Fr$ = 0.01. In the cascading regime, the surface particles assume the expected S-curved shape. At $Fr$ = 0.5, the flow enters the cataracting regime, where particles rise to a steep angle along the drum wall before detaching and falling back to the bottom of the drum. Finally, by $Fr$ = 1.5, the flow has transitioned into the centrifuging regime. At high Froude numbers, particles are thrown against the inner wall of the drum and rotate at the same velocity as the container. The observed flow patterns match the predicted motion and transition behaviors described in \citet{mellmann2001} and \citet{henein1983modeling}. 

\subsubsection{Angle of repose}
\label{sec:drum_results_repose}

\begin{figure*}
\begin{centering}
\includegraphics{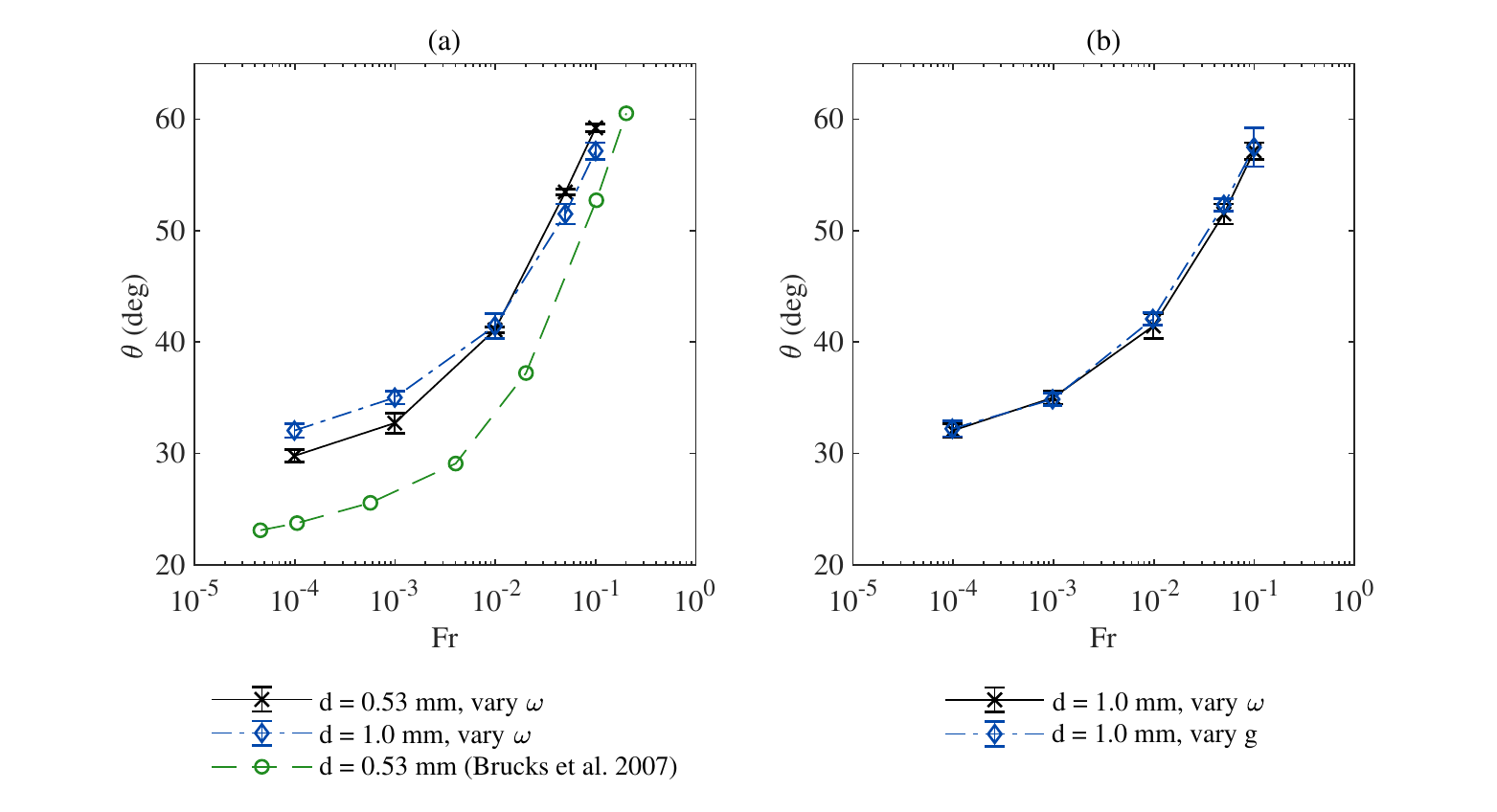}
\caption{Dynamic angle of repose $\theta$ for tumbler simulations with different Froude numbers. Plot (a) compares simulation results against data from \citet{brucks2007}. Plot (b) shows simulation results when the Froude number is obtained by varying either drum rotation speed or gravity-level (see Eq. \ref{eq:drum_fr}). In one test set, $\omega$ varies from 1.7 to 55 rpm. In the other, $g$ varies from $0.005g_o$ to $5g_o$.}
\label{fig:drum_repose}  
\end{centering}
\end{figure*}

Simulations with 0.53 mm particles and a 60 mm drum were carried out to provide a full-scale comparison with the \citet{brucks2007} experiments. The simulations cover a range of Froude numbers by holding gravity-level constant at 1$g_o$ while varying drum velocity from 1.7 to 212 rpm. The trend of increasing $\theta$ with $Fr$ shown in Fig. \ref{fig:drum_repose} matches experimental data, but the magnitudes of the repose angles are on the order of 5 to 7 degrees higher than observed in the physical tests. One explanation for the discrepancy could be a mismatch in material properties between the real and simulated beads. Previous studies have found that sliding, rolling, and wall friction have the biggest influences on tumbler flow behavior, while Young's modulus, Poisson's ratio, and coefficient of restitution are less important, given that the values fall within reasonable ranges \citep{qi2015, yan2015, chou2016}. Another explanation for the discrepancy could be that the particles fixed along the inside of the drums walls result in a more influential boundary condition than created by the walls of the experimental drum, which were lined with 60-grit sandpaper. 

Each full scale simulation takes approximately 4,000 cpu hours or 5 days on a single processor to complete. To reduce computation time, all remaining simulations were conducted using 1 mm diameter particles. Increasing particle size while keeping drum diameter fixed reduces the number of particles in the system from about 64,500 to 9,200. A comparison between the repose angles for the two different particle sizes can be found in Fig. \ref{fig:drum_repose}. Experiments have shown that the repose angle either decreases or remains constant when drum to particle diameter (D/d) increases, at least for low rotational velocities \citep{liu2005, brucks2007, dury1998}. A similar phenomenon occurs when the ratio of drum length to particle diameter (L/d) increases \citep{dury1998, yang2008}. Consistent with these findings, the simulations with the 1.0 mm particles reach higher repose angles than those with the 0.53 mm particle when $Fr < 0.01$. When $Fr > 0.01$, the trend changes, and higher repose angles are observed for the smaller particles. \citet{dury1998} reported the same outcome when investigating the effects of boundary conditions on repose angle. At lower Froude numbers, particles are more densely packed and lose the bulk of their energy through frequently occurring collisions. As Froude number increases however, the particle bed dilates and collision frequency decreases \citep{yang2008}. Extrapolating from this line of thought, it is possible that friction and boundary conditions are more influential at lower Froude numbers, where the particles are more constrained and inter-particle interactions dominate the flow.

The above tests cover a range of drum rotation speeds, but only one gravity level. Since $\theta$ is dependent on both $\omega$ and $g$, varying gravity instead of rotation speed should produce the same results. To verify, more simulations were executed with drum rotation speed fixed at 3.8 rpm and gravity-level ranging from $3.2 \times 10^{-4}g_o$ to $5g_o$. As expected from the \citet{brucks2007} experiments, the repose angles collapse onto a single curve (see Fig. \ref{fig:drum_repose}).  

\subsubsection{Velocity profile in the rolling regime}
\label{sec:drum_results_velocity}

\begin{figure*}
\begin{centering}
\includegraphics{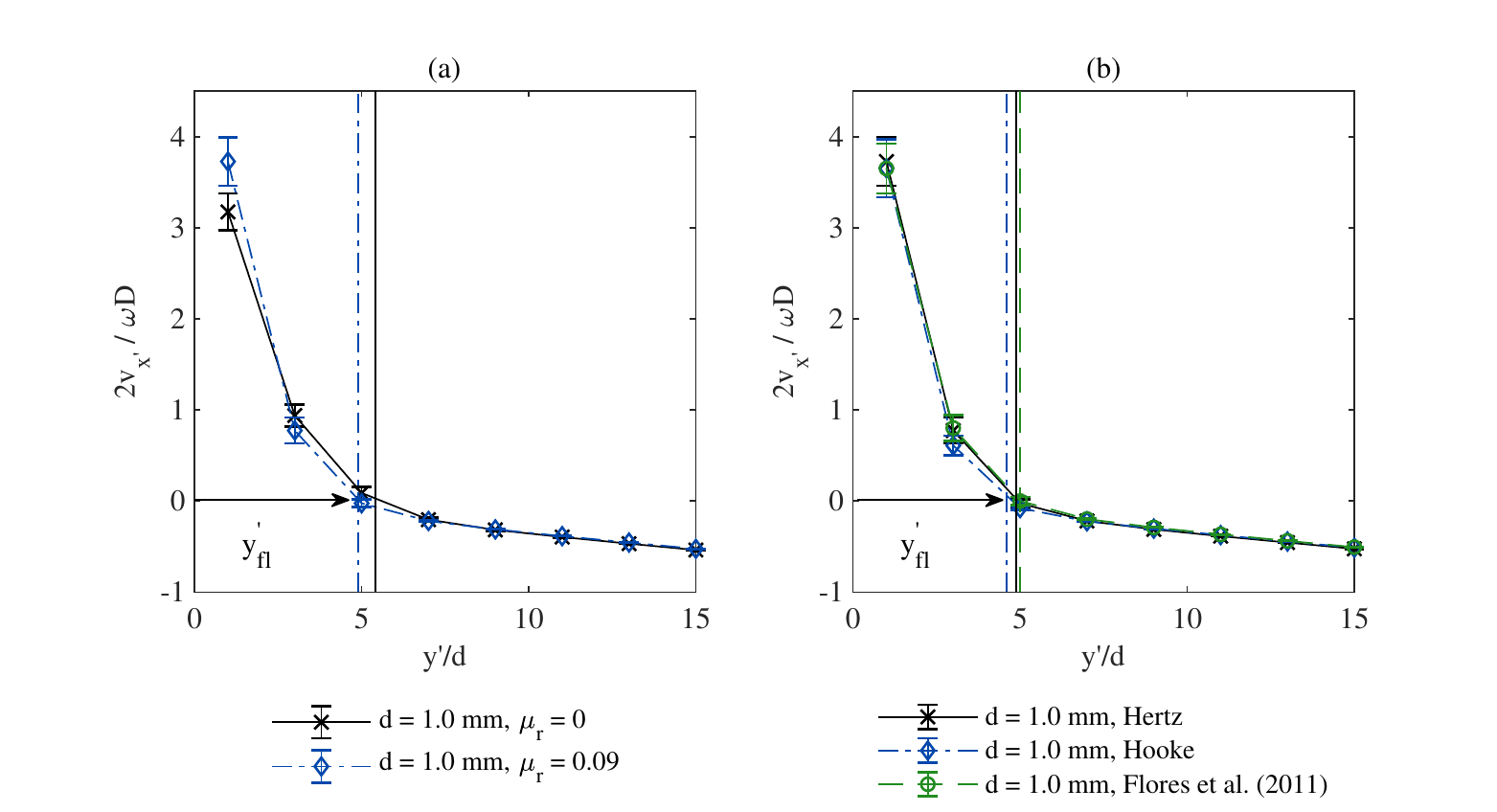}
\caption{Average streamwise velocity $v_x^\prime$ as a function of distance from the free surface for tumbler simulations where d = 1.0 mm. The vertical lines on the plot depict flowing layer thickness $y^\prime_{fl}$. The streamwise velocity is normalized by drum rotation speed $\omega$ and drum diameter $D$, and the distance from the free surface is normalized by particle diameter $d$. Plot (a) shows the velocity profiles for tests with the Hertz force model and different rolling friction coefficients and plot (b) compares results for different force models when $\mu_r$ = 0.09. The higher rolling friction leads to increased velocity at the free surface while the different force models show little to no variation in the velocity profile.}
\label{fig:drum_vprofile}
\end{centering}
\end{figure*}

Rolling friction is varied in order to understand the influence of $\mu_r$ on the simulation results. Tests are conducted within the rolling regime, at Fr = $1\times 10^{-3}$, so that streamwise velocities can be compared in addition to repose angles. The tests show that $\theta$ increases more than 5 degrees when $\mu_r$ changes from 0 to 0.09 (see Table \ref{tab:drum_results}). The higher angles produce more energetic particles, increasing the average velocity on the bed's surface (see Fig. \ref{fig:drum_vprofile}). In Table \ref{tab:drum_results}, an increase in $\theta$ typically coincides with an increase in flowing layer thickness. However, $y^\prime_{fl}$ is actually lower when $\mu_r$ = 0.09 than when $\mu_r$ = 0. This is because energy dissipates more quickly through the bed when the total contact torque takes into account a rolling resistance moment. Rolling friction increases the rate of velocity change through the flowing layer and reduces the flowing layer thickness. These results match findings from \citet{chou2016}, who conducts a detailed investigation into the effects of friction on tumbler flow.

The Hooke, Hertz, and \citet{flores2011} force models are also compared at $Fr = 1\times 10^{-3}$ and $\mu_r$ = 0.09. The streamwise velocity profiles for the three models are similar, suggesting that the non-physical behavior associated with the Hookean and Hertzian models has little impact on the bulk response of the system (see Fig. \ref{fig:drum_vprofile}). Additional testing is required to determine if this observation holds across different applications and flow-states. 

\subsubsection{Flow behavior with cohesion}
\label{sec:drum_results_cohesion}

\begin{figure*}
\begin{centering}
\includegraphics{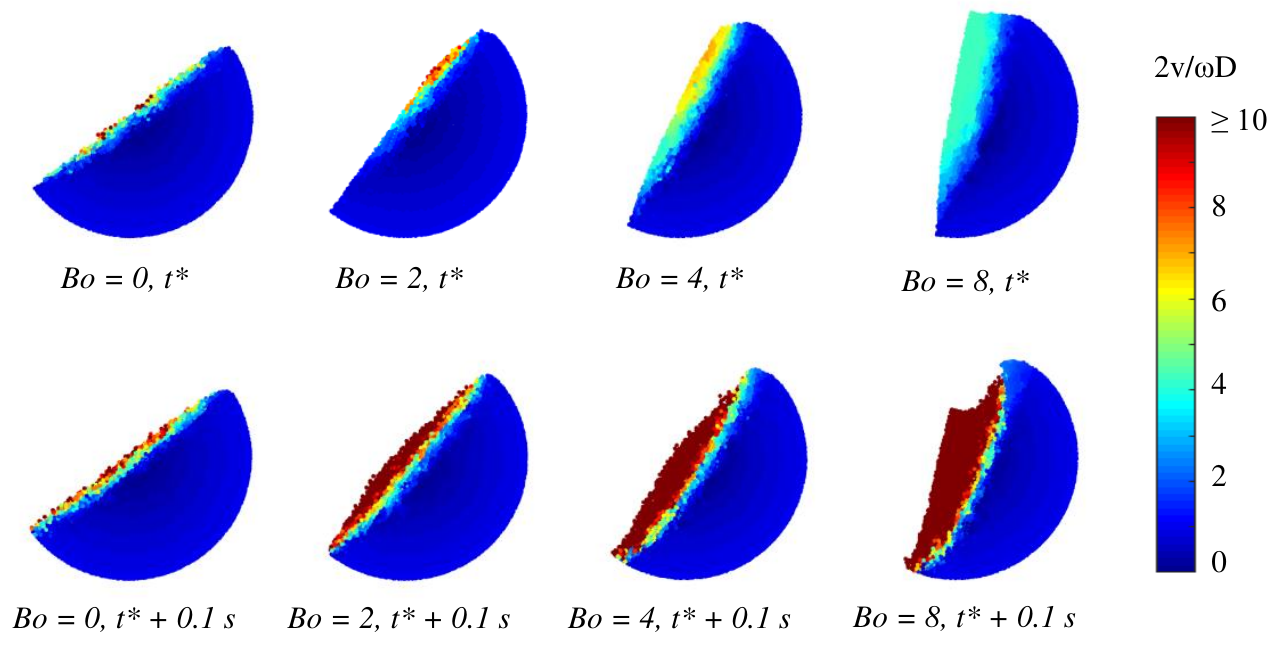}
\caption{Flow patterns for simulation with different granular Bond numbers when $d$ = 1 mm, $\mu_r$ = 0.09, $g = 0.5g_o$, $\omega$ = 3.8 rpm, and $Fr$ = 0.001. Particles are colored by velocity magnitude, normalized by drum diameter D and drum rotation speed $\omega$. $t^*$ is the time where the system reaches a maximum stable angle before experiencing its first avalanche.}
\label{fig:drum_vel_cohesion}
\end{centering}
\end{figure*}

The tests presented in Sections \ref{sec:drum_results_behavior} – \ref{sec:drum_results_velocity} neglect inter-particle cohesion. Here, we use the \citet{perko2001} cohesion model to explore how cohesion influences flow behavior and dynamic angle of repose for simulations under both Earth-gravity and reduced-gravity levels. First, we vary cohesion while leaving the gravity-level constant at $0.5g_o$ and the drum rotation speed constant at 3.8 rpm. The Froude number for the given configuration is 0.001, and the cohesion multiplier $C_3$ is selected such that the granular Bond number ranges from 0.5 to 8 (see Table \ref{tab:drum_results}). Fig. \ref{fig:drum_vel_cohesion} illustrates the flow patterns and the normalized flow velocities for tests where $Bo$ = 0, 2, 4, and 8. The top row of the figure corresponds to the time $t$ where the system reaches a maximum stable angle before experiencing its first avalanche ($t = t^*$) and the bottom row shows the state of the system 0.1 seconds later ($t = t^* + 0.1$ s). The time difference between the top and bottom images corresponds to a small angular distance and was selected simply to illustrate the material's transition from a semi-solid to a flowing state.

The flow behavior when $Bo$ = 0 is consistent with the rolling regime, as evident from the thin, fast-moving layer on the surface of the bed. As the Bond number increases however, the flow undergoes several observable changes. First, the particles begin to avalanche in clusters rather than individually. The larger the bond number, the larger the collapsing cluster. Once the flow is initiated (i.e., at $t = t^* + 0.1$ s), the surface profile evolves from being flat or slightly concave at $Bo$ = 0 to convex at $Bo$ = 2 and 4 and to irregularly shaped at $Bo$ = 8. Additionally, the thickness of the high-velocity flowing region on the surface of the bed increases as $Bo$ increases.

\begin{figure*}
\begin{centering}
\includegraphics{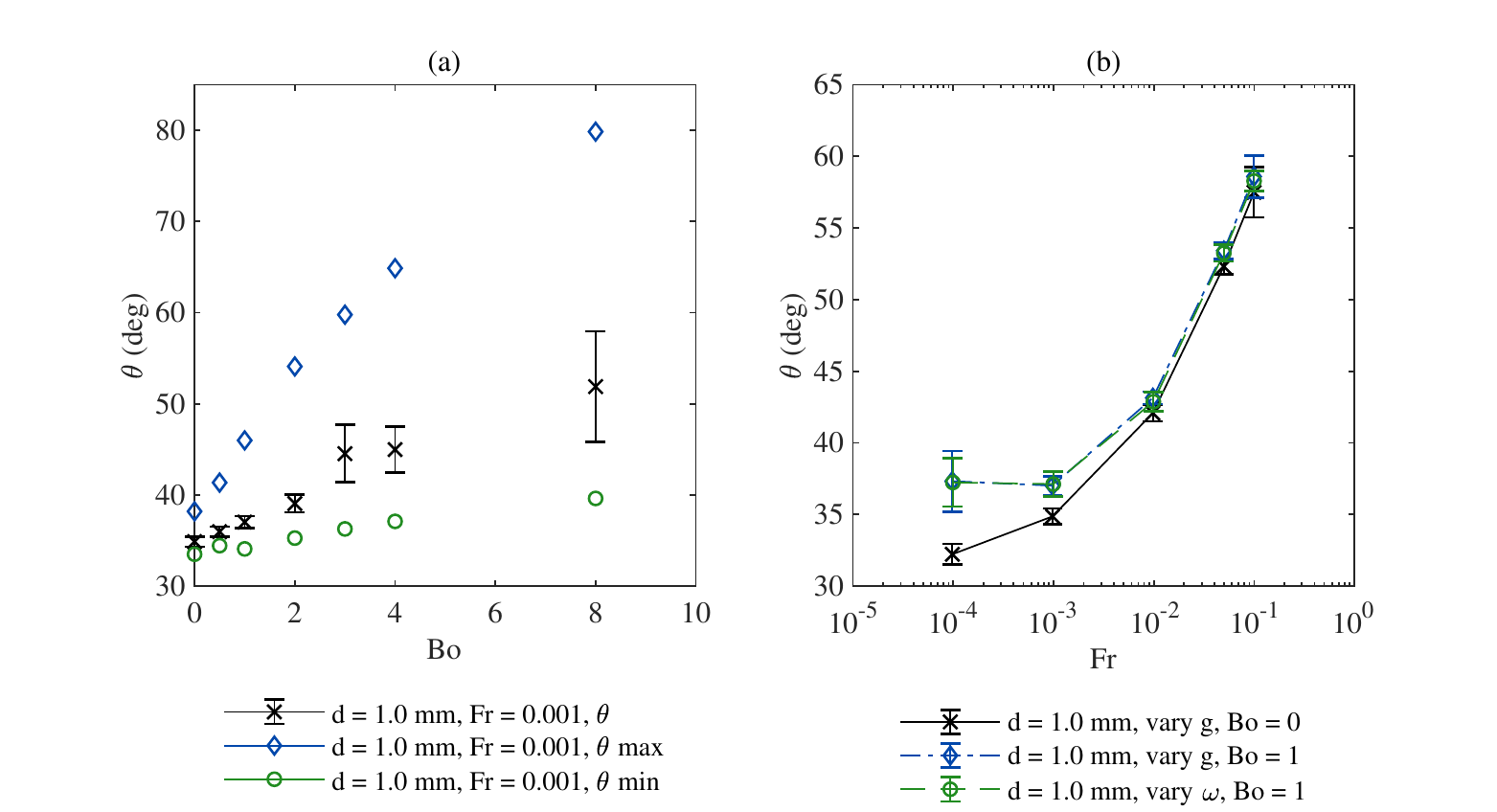}
\caption{Dynamic angle of repose $\theta$ for tumbler simulations with different Froude and Bond numbers. The error bars represent the standard deviation of the mean $\theta$ measurement over 1 real second of stead-state flow, while $\theta$ max and $\theta$ min are the before and after angle measurements for the flow's first avalanche. Plot (a) compares simulations with different Bond numbers when $g = 0.5g_o$ and $\omega$ = 3.8 rpm. Plot (b) shows simulation results when Froude number is obtained by varying either drum rotation speed or gravity-level. In one test set, $\omega$ varies from 1.7 to 55 rpm. In the others, $g$ varies from $0.005g_o$ to $5g_o$.}
\label{fig:drum_theta_cohesion}
\end{centering}
\end{figure*}

Shortly after flow initiates, the behavior reaches a steady state. In Fig. \ref{fig:drum_theta_cohesion} (a), we plot the dynamic angle of repose as a function of the Bond number. As expected, both the dynamic angle of repose and the maximum stable angle of repose increase as the Bond number increases. The error bars on the angle measurements are large when $Bo$ = 2, 3, and 4, indicating that the system experiences periodic-avalanching at higher cohesion values. Observations regarding the collapse/avalanching behavior, the surface profiles, the angle of repose, and the flowing-layer velocity are qualitatively consistent with findings from previous experiments and simulations performed at $1g_o$ \citep{ nase2001, walton2007, brewster2009, chou2011, liu2013, jarray2017, jarray2019}.

In the next set of tests, we check if the relationship between the dynamic angle of repose and the Froude number holds when inter-particle cohesion is non-zero. Like in Section \ref{sec:drum_results_repose}, the Froude number is controlled by varying either the drum rotation speed or the gravity-level. For test cases where $Bo$ = 1 and $g = g_o$, $\omega$ ranges from 1.7 to 55 rpm and the cohesion multiplier $C_3$ remains constant at 51.37 g s\textsuperscript{-2}. For test cases where $Bo$ = 1 and $\omega = 3.8$ rpm, the gravity-level ranges from $0.005g_o$ to $5g_o$ and $C_3$ varies from 0.252 to 256.8 g s\textsuperscript{-2} (see Table \ref{tab:drum_results}). 

Fig. \ref{fig:drum_theta_cohesion} (b) shows the dynamic angle of repose plotted as a function of the Froude number for the second set of simulations. The data collapses onto a single curve for all cases where $Bo$ = 1, just as it did for the cohesionless system (see Fig. \ref{fig:drum_repose}). In Fig. \ref{fig:drum_theta_cohesion} (b), we also see that the dynamic angle of repose is slightly higher when $Bo$ = 1 than when $Bo$ = 0, though the difference is more pronounced at lower Froude numbers. Based on the results from the first cohesion test and a study by \citet{walton2007}, we would expect the angle gap between the cohesionless and the cohesion-dominated system to grow as $Bo$ increases until some critical Bond number is reached. Above that critical number, the material would stop flowing and would fall apart in clumps or would simply rotate as a solid body. 

\citet{nase2001} conducted a piling, a hopper flow, and a tumbler study with wet granular material and controllable levels of capillary cohesion. The authors found that the static angle of repose and the discharge rate for the piling and the hopper tests change drastically as soon as the Bond number exceeds a $Bo$ = 1 threshold. However, their experimental data shows that the dynamic angle of repose for the tumbler tests does not jump or change dramatically when $Bo \geq 1$. Consequently, \citet{nase2001} cannot distinguish a clear transition between the flowing and cohesive states in the tumbler using the Bond number alone as a characterization tool. Like \citet{nase2001}, our tests show a gradual increase in $\theta$ as $Bo$ increases. This suggests that in addition to gravitational and cohesive forces, shearing and collision forces play non-negligible roles in certain granular phenomena. Much more work is required to understand impacts of cohesion on different types of flows. 

\section{Conclusions}

The soft-sphere DEM code in \textsc{Chrono::Parallel} 4.0.0 was modified to include the \citet{flores2011} force model, the \citet{perko2001} cohesion model, and to account for rolling and spinning friction. These enhancements are relevant for both terrestrial and planetary science applications and are publicly available as of \textsc{Chrono} version 5.0.0. The code changes were validated using a combination of two-body and multi-body benchmarking tests. The two-body tests reveal that the normal, tangential, and cohesive force calculations are correctly implemented in the code, and that the sliding, rolling, and spinning models yield the expected behaviors. 

In Section \ref{sec:pile}, we compared experimental and numerical results for a `sand piling' test using 1 mm glass beads. We varied the coefficient of rolling friction in the simulations, and found that as expected, the pile's angle of repose increases as $\mu_r$ increases. The angle of repose given by the simulations best matches the experimental data when $\mu_r = 0.09$.

Finally, in Section \ref{sec:drum}, we present the results for the rotating drum simulations. We varied, among other parameters, gravity-level and cohesion, noting that solid bodies in our Solar System, from asteroids to planets, cover a wide range of gravity conditions. Overall, the simulation results match findings from other experimental and numerical works. We observe that the drum flow spans the rolling, cascading, cataracting, and centrifuging regimes when the Froude number $Fr$ increases from $1\times10^{-4}$ to 1.5 (see Fig. \ref{fig:drum_vel}). The regime transitions occur at the expected value of $Fr$, regardless of how $Fr$ is controlled (i.e., by changing rotational velocity or by changing gravity-level). The angle of repose and the flowing layer thickness were measured when $Fr$ $\leq$ 0.1. When all parameters aside from $Fr$ are held constant, $\theta$ and $y^\prime_{fl}$ increase with $Fr$. Otherwise, subtle differences are observed when particle size, rolling friction coefficient, and force model are varied. Flow patterns and regime transitions change when cohesion is introduced into the system, and the dynamic angle of repose increases as the granular Bond number increases. The simulation results with cohesion are also in agreement with previous experimental works. 

The soft-sphere DEM model in \textsc{Chrono::Parallel} accurately replicates known granular flow behaviors, even for varied gravity and cohesion levels. As part of future work, this platform will be used to study regolith dynamics and lander-surface interactions. This upcoming work will aid with the interpretation of surface-regolith images sent by current and past missions (e.g. Hayabusa2, OSIRIS-REx) and will help prepare for future ones, like JAXA's MMX mission to Phobos and Deimos and ESA's Hera mission to the binary asteroid Didymos.


\section*{Acknowledgements}

We would like to thank Dan Negrut and Radu Serban at the University of Wisconsin-Madison and the \textsc{Chrono} developers for their support and collaboration. We would also like to acknowledge Jos\'e Andrade and the Computational Geomechanics group at the California Institute of Technology for allowing us to use their laboratory equipment and facilities, and Lennart Klar for assisting with early code development. This project uses HPC resources from CALMIP under grant allocation 2019-P19030, and is jointly funded by the Centre National d'Etudes Spatiales (CNES) and the Institut Sup\'erieur de l'A\'eronautique et de l'Espace (ISAE) under a PhD research grant. Naomi Murdoch, Stephen Schwartz, and Patrick Michel acknowledge funding support from the French space agency CNES. Patrick Michel and Stephen Schwartz acknowledge funding from Academies of Excellence: Complex systems and Space, environment, risk, and resilience, part of the IDEX JEDI of the Universit\'e C\^ote d'Azur. Stephen Schwartz acknowledges grant number 80NSSC18K0226 as part of the OSIRIS-REx Participating Scientist Program.


\bibliographystyle{mnras}
\bibliography{mnras} 

\begin{thebibliography}{}
\makeatletter
\relax
\def\mn@urlcharsother{\let\do\@makeother \do\$\do\&\do\#\do\^\do\_\do\%\do\~}
\def\mn@doi{\begingroup\mn@urlcharsother \@ifnextchar [ {\mn@doi@}
  {\mn@doi@[]}}
\def\mn@doi@[#1]#2{\def\@tempa{#1}\ifx\@tempa\@empty \href
  {http://dx.doi.org/#2} {doi:#2}\else \href {http://dx.doi.org/#2} {#1}\fi
  \endgroup}
\def\mn@eprint#1#2{\mn@eprint@#1:#2::\@nil}
\def\mn@eprint@arXiv#1{\href {http://arxiv.org/abs/#1} {{\tt arXiv:#1}}}
\def\mn@eprint@dblp#1{\href {http://dblp.uni-trier.de/rec/bibtex/#1.xml}
  {dblp:#1}}
\def\mn@eprint@#1:#2:#3:#4\@nil{\def\@tempa {#1}\def\@tempb {#2}\def\@tempc
  {#3}\ifx \@tempc \@empty \let \@tempc \@tempb \let \@tempb \@tempa \fi \ifx
  \@tempb \@empty \def\@tempb {arXiv}\fi \@ifundefined
  {mn@eprint@\@tempb}{\@tempb:\@tempc}{\expandafter \expandafter \csname
  mn@eprint@\@tempb\endcsname \expandafter{\@tempc}}}

\bibitem[\protect\citeauthoryear{Ai, Chen, Rotter  \& Ooi}{Ai
  et~al.}{2011}]{ai2011}
Ai J.,  Chen J.-F.,  Rotter J.~M.,   Ooi J.~Y.,  2011, Powder Technology, 206,
  269

\bibitem[\protect\citeauthoryear{Alizadeh, Bertrand  \& Chaouki}{Alizadeh
  et~al.}{2014}]{alizadeh2014}
Alizadeh E.,  Bertrand F.,   Chaouki J.,  2014, AIChE Journal, 60, 60

\bibitem[\protect\citeauthoryear{Amstock}{Amstock}{1997}]{amstock1997}
Amstock J.~S.,  1997, Handbook of glass in construction.
McGraw Hill Professional

\bibitem[\protect\citeauthoryear{Anand, Curtis, Wassgren, Hancock  \&
  Ketterhagen}{Anand et~al.}{2008}]{anand2008}
Anand A.,  Curtis J.~S.,  Wassgren C.~R.,  Hancock B.~C.,   Ketterhagen W.~R.,
  2008, Chemical Engineering Science, 63, 5821

\bibitem[\protect\citeauthoryear{Asmar, Langston, Matchett  \& Walters}{Asmar
  et~al.}{2002}]{asmar2002}
Asmar B.,  Langston P.,  Matchett A.,   Walters J.,  2002, Computers \&
  chemical engineering, 26, 785

\bibitem[\protect\citeauthoryear{Beatini, Royer-Carfagni  \& Tasora}{Beatini
  et~al.}{2017}]{beatini2017}
Beatini V.,  Royer-Carfagni G.,   Tasora A.,  2017, Computers \& Structures,
  187, 88

\bibitem[\protect\citeauthoryear{Beverloo, Leniger  \& Van~de Velde}{Beverloo
  et~al.}{1961}]{beverloo1961}
Beverloo W.~A.,  Leniger H.~A.,   Van~de Velde J.,  1961, Chemical engineering
  science, 15, 260

\bibitem[\protect\citeauthoryear{Biele et~al.,}{Biele et~al.}{2015}]{biele2015}
Biele J.,  et~al., 2015, Science, 349, aaa9816

\bibitem[\protect\citeauthoryear{Bolz}{Bolz}{2019}]{bolz2019}
Bolz R.~E.,  2019, CRC handbook of tables for applied engineering science.
CRC press

\bibitem[\protect\citeauthoryear{Brewster, Grest  \& Levine}{Brewster
  et~al.}{2009}]{brewster2009}
Brewster R.,  Grest G.~S.,   Levine A.~J.,  2009, Physical Review E, 79, 011305

\bibitem[\protect\citeauthoryear{Brilliantov \& P{\"o}schel}{Brilliantov \&
  P{\"o}schel}{1998}]{brilliantov1998}
Brilliantov N.~V.,  P{\"o}schel T.,  1998, EPL (Europhysics Letters), 42, 511

\bibitem[\protect\citeauthoryear{Brisset, Colwell, Dove, Abukhalil, Cox  \&
  Mohammed}{Brisset et~al.}{2018}]{brisset2018}
Brisset J.,  Colwell J.,  Dove A.,  Abukhalil S.,  Cox C.,   Mohammed N.,
  2018, Progress in Earth and Planetary Science, 5, 73

\bibitem[\protect\citeauthoryear{Brown \& Richards}{Brown \&
  Richards}{1965}]{brown1965}
Brown R.,  Richards J.,  1965, Rheologica Acta, 4, 153

\bibitem[\protect\citeauthoryear{Brucks, Arndt, Ottino  \& Lueptow}{Brucks
  et~al.}{2007}]{brucks2007}
Brucks A.,  Arndt T.,  Ottino J.~M.,   Lueptow R.~M.,  2007, Physical Review E,
  75, 032301

\bibitem[\protect\citeauthoryear{Chen, Liu, Zhao, Xiao  \& Liu}{Chen
  et~al.}{2015}]{chen2015}
Chen H.,  Liu Y.,  Zhao X.,  Xiao Y.,   Liu Y.,  2015, Powder technology, 283,
  607

\bibitem[\protect\citeauthoryear{Chen, Xiao, Liu  \& Shi}{Chen
  et~al.}{2017}]{chen2017}
Chen H.,  Xiao Y.,  Liu Y.,   Shi Y.,  2017, Powder technology, 318, 507

\bibitem[\protect\citeauthoryear{Cheng, Santo, Heeres, Landshof, Farquhar, Gold
   \& Lee}{Cheng et~al.}{1997}]{cheng1997}
Cheng A.~F.,  Santo A.,  Heeres K.,  Landshof J.,  Farquhar R.,  Gold R.,   Lee
  S.,  1997, Journal of Geophysical Research: Planets, 102, 23695

\bibitem[\protect\citeauthoryear{{Cheng} et~al.,}{{Cheng}
  et~al.}{2017}]{cheng2017}
{Cheng} A.~F.,  et~al., 2017, in Lunar and Planetary Science Conference. Lunar
  and Planetary Science Conference.
p.~1510

\bibitem[\protect\citeauthoryear{Chou \& Hsiau}{Chou \& Hsiau}{2011}]{chou2011}
Chou S.,  Hsiau S.,  2011, Powder technology, 214, 491

\bibitem[\protect\citeauthoryear{Chou, Hu  \& Hsiau}{Chou
  et~al.}{2016}]{chou2016}
Chou S.,  Hu H.,   Hsiau S.,  2016, Advanced Powder Technology, 27, 1912

\bibitem[\protect\citeauthoryear{Co{\"\i}sson, Ferrari, Ferretti  \&
  Rozzi}{Co{\"\i}sson et~al.}{2016}]{coisson2016}
Co{\"\i}sson E.,  Ferrari L.,  Ferretti D.,   Rozzi M.,  2016, Procedia
  engineering, 161, 451

\bibitem[\protect\citeauthoryear{Colwell \& Taylor}{Colwell \&
  Taylor}{1999}]{colwell1999}
Colwell J.~E.,  Taylor M.,  1999, Icarus, 138, 241

\bibitem[\protect\citeauthoryear{Derjaguin, Muller  \& Toporov}{Derjaguin
  et~al.}{1975}]{derjaguin1975}
Derjaguin B.~V.,  Muller V.~M.,   Toporov Y.~P.,  1975, Journal of Colloid and
  interface science, 53, 314

\bibitem[\protect\citeauthoryear{Dury \& Ristow}{Dury \&
  Ristow}{1997}]{dury1997}
Dury C.~M.,  Ristow G.~H.,  1997, Journal de Physique I, 7, 737

\bibitem[\protect\citeauthoryear{Dury, Ristow, Moss  \& Nakagawa}{Dury
  et~al.}{1998}]{dury1998}
Dury C.~M.,  Ristow G.~H.,  Moss J.~L.,   Nakagawa M.,  1998, Physical Review
  E, 57, 4491

\bibitem[\protect\citeauthoryear{F{\'e}lix, Falk  \& D'Ortona}{F{\'e}lix
  et~al.}{2002}]{felix2002}
F{\'e}lix G.,  Falk V.,   D'Ortona U.,  2002, Powder Technology, 128, 314

\bibitem[\protect\citeauthoryear{Ferrari, Tasora, Masarati  \& Lavagna}{Ferrari
  et~al.}{2017}]{ferrari2017}
Ferrari F.,  Tasora A.,  Masarati P.,   Lavagna M.,  2017, Multibody System
  Dynamics, 39, 3

\bibitem[\protect\citeauthoryear{Fleischmann, Serban, Negrut  \&
  Jayakumar}{Fleischmann et~al.}{2016}]{fleischmann2016}
Fleischmann J.,  Serban R.,  Negrut D.,   Jayakumar P.,  2016, Journal of
  Computational and Nonlinear Dynamics, 11, 044502

\bibitem[\protect\citeauthoryear{Flores, Machado, Silva  \& Martins}{Flores
  et~al.}{2011}]{flores2011}
Flores P.,  Machado M.,  Silva M.~T.,   Martins J.~M.,  2011, Multibody system
  dynamics, 25, 357

\bibitem[\protect\citeauthoryear{Foerster, Louge, Chang  \& Allia}{Foerster
  et~al.}{1994}]{foerster1994}
Foerster S.~F.,  Louge M.~Y.,  Chang H.,   Allia K.,  1994, Physics of Fluids,
  6, 1108

\bibitem[\protect\citeauthoryear{Fujiwara et~al.,}{Fujiwara
  et~al.}{2006}]{fujiwara2006}
Fujiwara A.,  et~al., 2006, Science, 312, 1330

\bibitem[\protect\citeauthoryear{Glassmeier, Boehnhardt, Koschny, K{\"u}hrt  \&
  Richter}{Glassmeier et~al.}{2007}]{glassmeier2007}
Glassmeier K.-H.,  Boehnhardt H.,  Koschny D.,  K{\"u}hrt E.,   Richter I.,
  2007, Space Science Reviews, 128, 1

\bibitem[\protect\citeauthoryear{Gray \& Thornton}{Gray \&
  Thornton}{2005}]{gray2005}
Gray J.,  Thornton A.,  2005, Proceedings of the Royal Society A: Mathematical,
  Physical and Engineering Sciences, 461, 1447

\bibitem[\protect\citeauthoryear{Hartzell, Wang, Scheeres  \&
  Hor{\'a}nyi}{Hartzell et~al.}{2013}]{hartzell2013}
Hartzell C.,  Wang X.,  Scheeres D.,   Hor{\'a}nyi M.,  2013, Geophysical
  research letters, 40, 1038

\bibitem[\protect\citeauthoryear{Hartzell, Farrell  \& Marshall}{Hartzell
  et~al.}{2018}]{hartzell2018}
Hartzell C.~M.,  Farrell W.,   Marshall J.,  2018, Advances in Space Research,
  62, 2213

\bibitem[\protect\citeauthoryear{Henein, Brimacombe  \& Watkinson}{Henein
  et~al.}{1983a}]{henein1983experimental}
Henein H.,  Brimacombe J.,   Watkinson A.,  1983a, Metallurgical transactions
  B, 14, 191

\bibitem[\protect\citeauthoryear{Henein, Brimacombe  \& Watkinson}{Henein
  et~al.}{1983b}]{henein1983modeling}
Henein H.,  Brimacombe J.,   Watkinson A.,  1983b, Metallurgical Transactions
  B, 14, 207

\bibitem[\protect\citeauthoryear{Hofmann, Sierks  \& Blum}{Hofmann
  et~al.}{2017}]{hofmann2017}
Hofmann M.,  Sierks H.,   Blum J.,  2017, Monthly Notices of the Royal
  Astronomical Society, 469, S73

\bibitem[\protect\citeauthoryear{Hu, Liu  \& Wu}{Hu et~al.}{2018}]{hu2018}
Hu Z.,  Liu X.,   Wu W.,  2018, Powder technology, 340, 563

\bibitem[\protect\citeauthoryear{Huang, Nydal  \& Yao}{Huang
  et~al.}{2014}]{huang2014}
Huang Y.~J.,  Nydal O.~J.,   Yao B.,  2014, Powder technology, 253, 80

\bibitem[\protect\citeauthoryear{Iwashita \& Oda}{Iwashita \&
  Oda}{1998}]{iwashita1998}
Iwashita K.,  Oda M.,  1998, Journal of engineering mechanics, 124, 285

\bibitem[\protect\citeauthoryear{Jarray, Magnanimo, Ramaioli  \& Luding}{Jarray
  et~al.}{2017}]{jarray2017}
Jarray A.,  Magnanimo V.,  Ramaioli M.,   Luding S.,  2017, in EPJ Web of
  Conferences. p. 03078

\bibitem[\protect\citeauthoryear{Jarray, Magnanimo  \& Luding}{Jarray
  et~al.}{2019}]{jarray2019}
Jarray A.,  Magnanimo V.,   Luding S.,  2019, Powder technology, 341, 126

\bibitem[\protect\citeauthoryear{Jaumann et~al.,}{Jaumann
  et~al.}{2019}]{jaumann2019}
Jaumann R.,  et~al., 2019, Science, 365, 817

\bibitem[\protect\citeauthoryear{Jiang, Yu  \& Harris}{Jiang
  et~al.}{2005}]{jiang2005}
Jiang M.,  Yu H.-S.,   Harris D.,  2005, Computers and Geotechnics, 32, 340

\bibitem[\protect\citeauthoryear{Kharaz, Gorham  \& Salman}{Kharaz
  et~al.}{2001}]{kharaz2001}
Kharaz A.,  Gorham D.,   Salman A.,  2001, Powder Technology, 120, 281

\bibitem[\protect\citeauthoryear{Kleinhans, Markies, De~Vet, Postema
  et~al.}{Kleinhans et~al.}{2011}]{kleinhans2011}
Kleinhans M.,  Markies H.,  De~Vet S.,  Postema F.,   et~al., 2011, Journal of
  Geophysical Research: Planets, 116

\bibitem[\protect\citeauthoryear{Kruggel-Emden, Simsek, Rickelt, Wirtz  \&
  Scherer}{Kruggel-Emden et~al.}{2007}]{kruggel2007}
Kruggel-Emden H.,  Simsek E.,  Rickelt S.,  Wirtz S.,   Scherer V.,  2007,
  Powder Technology, 171, 157

\bibitem[\protect\citeauthoryear{{Kuramoto}, {Kawakatsu}  \&
  {Fujimoto}}{{Kuramoto} et~al.}{2018}]{kuramoto2018}
{Kuramoto} K.,  {Kawakatsu} Y.,   {Fujimoto} M.,  2018, in European Planetary
  Science Congress. pp EPSC2018--1036

\bibitem[\protect\citeauthoryear{Lauretta et~al.,}{Lauretta
  et~al.}{2017}]{lauretta2017}
Lauretta D.,  et~al., 2017, Space Science Reviews, 212, 925

\bibitem[\protect\citeauthoryear{Li, Xu  \& Thornton}{Li et~al.}{2005}]{li2005}
Li Y.,  Xu Y.,   Thornton C.,  2005, Powder Technology, 160, 219

\bibitem[\protect\citeauthoryear{Liu, Specht  \& Mellmann}{Liu
  et~al.}{2005}]{liu2005}
Liu X.~Y.,  Specht E.,   Mellmann J.,  2005, Powder Technology, 154, 125

\bibitem[\protect\citeauthoryear{Liu, Yang  \& Yu}{Liu et~al.}{2013}]{liu2013}
Liu P.,  Yang R.,   Yu A.,  2013, Chemical Engineering Science, 86, 99

\bibitem[\protect\citeauthoryear{Luding}{Luding}{2008}]{luding2008}
Luding S.,  2008, Granular matter, 10, 235

\bibitem[\protect\citeauthoryear{Maurel, Michel, Biele, Ballouz  \&
  Thuillet}{Maurel et~al.}{2018}]{maurel2018}
Maurel C.,  Michel P.,  Biele J.,  Ballouz R.-L.,   Thuillet F.,  2018,
  Advances in Space Research, 62, 2099

\bibitem[\protect\citeauthoryear{Mazhar, Heyn, Pazouki, Melanz, Seidl,
  Bartholomew, Tasora  \& Negrut}{Mazhar et~al.}{2013}]{mazhar2013}
Mazhar H.,  Heyn T.,  Pazouki A.,  Melanz D.,  Seidl A.,  Bartholomew A.,
  Tasora A.,   Negrut D.,  2013, Mechanical Sciences, 4, 49

\bibitem[\protect\citeauthoryear{Mazhar, Osswald  \& Negrut}{Mazhar
  et~al.}{2016}]{mazhar2016}
Mazhar H.,  Osswald T.,   Negrut D.,  2016, Additive Manufacturing, 12, 291

\bibitem[\protect\citeauthoryear{Mellmann}{Mellmann}{2001}]{mellmann2001}
Mellmann J.,  2001, Powder technology, 118, 251

\bibitem[\protect\citeauthoryear{{Michel} et~al.,}{{Michel}
  et~al.}{2018}]{michel2018}
{Michel} P.,  et~al., 2018, Advances in Space Research, 62, 2261

\bibitem[\protect\citeauthoryear{Mohamed \& Gutierrez}{Mohamed \&
  Gutierrez}{2010}]{mohamed2010}
Mohamed A.,  Gutierrez M.,  2010, Granular Matter, 12, 527

\bibitem[\protect\citeauthoryear{{Murdoch}, {S{\'a}nchez}, {Schwartz}  \&
  {Miyamoto}}{{Murdoch} et~al.}{2015}]{murdoch2015}
{Murdoch} N.,  {S{\'a}nchez} P.,  {Schwartz} S.~R.,   {Miyamoto} H.,  2015,
  {Asteroid Surface Geophysics}.
The University of Arizona Press, pp 767--792

\bibitem[\protect\citeauthoryear{Murdoch, Avila~Martinez, Sunday, Zenou,
  Cherrier, Cadu  \& Gourinat}{Murdoch et~al.}{2017}]{murdoch2017}
Murdoch N.,  Avila~Martinez I.,  Sunday C.,  Zenou E.,  Cherrier O.,  Cadu A.,
   Gourinat Y.,  2017, Monthly Notices of the Royal Astronomical Society, 468,
  1259

\bibitem[\protect\citeauthoryear{Myers \& Sellers}{Myers \&
  Sellers}{1971}]{myers1971}
Myers M.,  Sellers M.,  1971, Research Project Report, University of Cambridge

\bibitem[\protect\citeauthoryear{Nakashima, Shioji, Kobayashi, Aoki, Shimizu,
  Miyasaka  \& Ohdoi}{Nakashima et~al.}{2011}]{nakashima2011}
Nakashima H.,  Shioji Y.,  Kobayashi T.,  Aoki S.,  Shimizu H.,  Miyasaka J.,
  Ohdoi K.,  2011, Journal of terramechanics, 48, 17

\bibitem[\protect\citeauthoryear{Nase, Vargas, Abatan  \& McCarthy}{Nase
  et~al.}{2001}]{nase2001}
Nase S.~T.,  Vargas W.~L.,  Abatan A.~A.,   McCarthy J.,  2001, Powder
  Technology, 116, 214

\bibitem[\protect\citeauthoryear{Pazouki, Kwarta, Williams, Likos, Serban,
  Jayakumar  \& Negrut}{Pazouki et~al.}{2017}]{pazouki2017}
Pazouki A.,  Kwarta M.,  Williams K.,  Likos W.,  Serban R.,  Jayakumar P.,
  Negrut D.,  2017, Physical Review E, 96, 042905

\bibitem[\protect\citeauthoryear{Perko, Nelson  \& Sadeh}{Perko
  et~al.}{2001}]{perko2001}
Perko H.~A.,  Nelson J.~D.,   Sadeh W.~Z.,  2001, Journal of geotechnical and
  geoenvironmental engineering, 127, 371

\bibitem[\protect\citeauthoryear{Qi, Xu, Zhou, Chen, Ge  \& Li}{Qi
  et~al.}{2015}]{qi2015}
Qi H.,  Xu J.,  Zhou G.,  Chen F.,  Ge W.,   Li J.,  2015, Particuology, 22,
  119

\bibitem[\protect\citeauthoryear{Richardson, Quinn, Stadel  \& Lake}{Richardson
  et~al.}{2000}]{richardson2000}
Richardson D.~C.,  Quinn T.,  Stadel J.,   Lake G.,  2000, Icarus, 143, 45

\bibitem[\protect\citeauthoryear{Richardson, Walsh, Murdoch  \&
  Michel}{Richardson et~al.}{2011}]{richardson2011}
Richardson D.~C.,  Walsh K.~J.,  Murdoch N.,   Michel P.,  2011, Icarus, 212,
  427

\bibitem[\protect\citeauthoryear{Russell et~al.,}{Russell
  et~al.}{2007}]{russell2007}
Russell C.,  et~al., 2007, Earth, Moon, and Planets, 101, 65

\bibitem[\protect\citeauthoryear{S{\'a}nchez \& Scheeres}{S{\'a}nchez \&
  Scheeres}{2012}]{sanchez2012}
S{\'a}nchez D.~P.,  Scheeres D.~J.,  2012, Icarus, 218, 876

\bibitem[\protect\citeauthoryear{S{\'a}nchez \& Scheeres}{S{\'a}nchez \&
  Scheeres}{2014}]{sanchez2014}
S{\'a}nchez P.,  Scheeres D.~J.,  2014, Meteoritics \& Planetary Science, 49,
  788

\bibitem[\protect\citeauthoryear{Santos, Barrozo, Duarte, Weigler  \&
  Mellmann}{Santos et~al.}{2016}]{santos2016}
Santos D.~A.,  Barrozo M.~A.,  Duarte C.~R.,  Weigler F.,   Mellmann J.,  2016,
  Advanced Powder Technology, 27, 692

\bibitem[\protect\citeauthoryear{Scheeres, Hartzell, S{\'a}nchez  \&
  Swift}{Scheeres et~al.}{2010}]{scheeres2010}
Scheeres D.~J.,  Hartzell C.~M.,  S{\'a}nchez P.,   Swift M.,  2010, Icarus,
  210, 968

\bibitem[\protect\citeauthoryear{Schwager \& P{\"o}schel}{Schwager \&
  P{\"o}schel}{2008}]{schwager2008}
Schwager T.,  P{\"o}schel T.,  2008, Physical Review E, 78, 051304

\bibitem[\protect\citeauthoryear{Schwartz, Richardson  \& Michel}{Schwartz
  et~al.}{2012}]{schwartz2012}
Schwartz S.~R.,  Richardson D.~C.,   Michel P.,  2012, Granular Matter, 14, 363

\bibitem[\protect\citeauthoryear{Serban, Taylor, Negrut  \& Tasora}{Serban
  et~al.}{2019}]{serban2019}
Serban R.,  Taylor M.,  Negrut D.,   Tasora A.,  2019, International Journal of
  Vehicle Performance, 5

\bibitem[\protect\citeauthoryear{Stadel}{Stadel}{2001}]{stadel2001}
Stadel J.~G.,  2001, PhDT, p.~3657

\bibitem[\protect\citeauthoryear{Sugita et~al.,}{Sugita
  et~al.}{2019}]{sugita2019}
Sugita S.,  et~al., 2019, Science, 364, eaaw0422

\bibitem[\protect\citeauthoryear{Tancredi, Maciel, Heredia, Richeri  \&
  Nesmachnow}{Tancredi et~al.}{2012}]{tancredi2012}
Tancredi G.,  Maciel A.,  Heredia L.,  Richeri P.,   Nesmachnow S.,  2012,
  Monthly Notices of the Royal Astronomical Society, 420, 3368

\bibitem[\protect\citeauthoryear{Tardivel, Lange  \& the MMX
  Rover~Team}{Tardivel et~al.}{2019}]{tardivel2019}
Tardivel S.,  Lange C.,   the MMX Rover~Team .,  2019, in the 13th Low Cost
  Planetary Missions conference.

\bibitem[\protect\citeauthoryear{Tasora et~al.,}{Tasora
  et~al.}{2016}]{tasora2015}
Tasora A.,  et~al., 2016, in High Performance Computing in Science and
  Engineering. Springer International Publishing, pp 19--49

\bibitem[\protect\citeauthoryear{Thuillet et~al.,}{Thuillet
  et~al.}{2018}]{thuillet2018}
Thuillet F.,  et~al., 2018, Astronomy \& Astrophysics, 615, A41

\bibitem[\protect\citeauthoryear{Tsuji, Tanaka  \& Ishida}{Tsuji
  et~al.}{1992}]{tsuji1992}
Tsuji Y.,  Tanaka T.,   Ishida T.,  1992, Powder technology, 71, 239

\bibitem[\protect\citeauthoryear{Ulamec, Michel, Grott, Bottger, Hubers,
  Murdoch  \& et al.}{Ulamec et~al.}{2020}]{ulamec2020}
Ulamec S.,  Michel P.,  Grott M.,  Bottger U.,  Hubers H.-W.,  Murdoch N.,   et
  al. 2020, Acta Astronautica, Submitted

\bibitem[\protect\citeauthoryear{Walton, De~Moor  \& Gill}{Walton
  et~al.}{2007}]{walton2007}
Walton O.~R.,  De~Moor C.~P.,   Gill K.~S.,  2007, Granular Matter, 9, 353

\bibitem[\protect\citeauthoryear{Wang, Schwan, Hsu, Gr{\"u}n  \&
  Hor{\'a}nyi}{Wang et~al.}{2016}]{wang2016}
Wang X.,  Schwan J.,  Hsu H.-W.,  Gr{\"u}n E.,   Hor{\'a}nyi M.,  2016,
  Geophysical Research Letters, 43, 6103

\bibitem[\protect\citeauthoryear{Watanabe et~al.,}{Watanabe
  et~al.}{2019}]{watanabe2019}
Watanabe S.,  et~al., 2019, Science, 364, 268

\bibitem[\protect\citeauthoryear{Wu, Thornton  \& Li}{Wu et~al.}{2003}]{wu2003}
Wu C.-Y.,  Thornton C.,   Li L.-Y.,  2003, Advanced Powder Technology, 14, 435

\bibitem[\protect\citeauthoryear{Xiang, Munjiza, Latham  \& Guises}{Xiang
  et~al.}{2009}]{xiang2009}
Xiang J.,  Munjiza A.,  Latham J.-P.,   Guises R.,  2009, Engineering
  Computations, 26, 673

\bibitem[\protect\citeauthoryear{Xu, Xu, Zhou, Du  \& Hu}{Xu
  et~al.}{2010}]{xu2010}
Xu Y.,  Xu C.,  Zhou Z.,  Du J.,   Hu D.,  2010, Particuology, 8, 141

\bibitem[\protect\citeauthoryear{Yan, Wilkinson, Stitt  \& Marigo}{Yan
  et~al.}{2015}]{yan2015}
Yan Z.,  Wilkinson S.,  Stitt E.,   Marigo M.,  2015, Computational Particle
  Mechanics, 2, 283

\bibitem[\protect\citeauthoryear{Yang, Yu, McElroy  \& Bao}{Yang
  et~al.}{2008}]{yang2008}
Yang R.,  Yu A.,  McElroy L.,   Bao J.,  2008, Powder Technology, 188, 170

\bibitem[\protect\citeauthoryear{Yu, Richardson, Michel, Schwartz  \&
  Ballouz}{Yu et~al.}{2014}]{yu2014}
Yu Y.,  Richardson D.~C.,  Michel P.,  Schwartz S.~R.,   Ballouz R.-L.,  2014,
  Icarus, 242, 82

\bibitem[\protect\citeauthoryear{Yu, Elghannay  \& Tafti}{Yu
  et~al.}{2017}]{yu2017}
Yu K.,  Elghannay H.~A.,   Tafti D.,  2017, Powder technology, 319, 102

\bibitem[\protect\citeauthoryear{Zhang \& Whiten}{Zhang \&
  Whiten}{1996}]{zhang1996}
Zhang D.,  Whiten W.,  1996, Powder technology, 88, 59

\bibitem[\protect\citeauthoryear{Zhang et~al.,}{Zhang et~al.}{2017}]{zhang2017}
Zhang Y.,  et~al., 2017, Icarus, 294, 98

\bibitem[\protect\citeauthoryear{Zhou, Wright, Yang, Xu  \& Yu}{Zhou
  et~al.}{1999}]{zhou1999}
Zhou Y.,  Wright B.,  Yang R.,  Xu B.~H.,   Yu A.-B.,  1999, Physica A:
  Statistical Mechanics and its Applications, 269, 536

\bibitem[\protect\citeauthoryear{Zhou, Xu, Yu  \& Zulli}{Zhou
  et~al.}{2002}]{zhou2002}
Zhou Y.,  Xu B.~H.,  Yu A.-B.,   Zulli P.,  2002, Powder technology, 125, 45

\makeatother
\end{thebibliography}


\appendix

\section{Stiffness and damping coefficients}
\label{app:kandg}

In \textsc{Chrono} 5.0.0, contact forces are calculated using one of three models: the Hooke model, the Hertz model, or the \citet{flores2011} model. Eqs.\ref{eq:app_hooke} - \ref{eq:app_flores} provide the full set of equations associated with each model. In the normal force equations, $\vec{F}_n$ is the normal force, $k_n$ is normal stiffness, $g_n$ is normal damping, $\vec{v}_n$ is the normal component of the relative velocity at the point of contact, $\delta_n$ is the normal overlap, and $\hat{n}$ is the unit vector pointing from one particle center to the other. In the tangential force equations, $\vec{F}_t$ is the tangential force, $k_t$ is tangential stiffness, $g_t$ is tangential damping, $\vec{v}_t$ is the tangential component of the relative velocity at the point of contact, $\vec{\delta}_t$ is the tangential displacement vector, and $e$ is the coefficient of restitution. In the stiffness and damping equations, $R$, $M$, $\bar{E}$ and $\bar{G}$ are respectively the effective radius, the effective mass, the effective Young's modulus, and the effective Shear modulus of the contacting pair. In Eqs. \ref{eq:app_R} - \ref{eq:app_G}, $r$, $m$, $E$, and $\nu$ are respectively the radius, the mass, the Young's modulus, and the Poisson's ratio of the individual particles in the colliding pair.

\begin{equation}
\label{eq:app_R}
    R = \left( \frac{1}{r_i} + \frac{1}{r_j} \right) ^{-1}
\end{equation}

\begin{equation}
\label{eq:app_M}
    M = \left( \frac{1}{m_i} + \frac{1}{m_j} \right) ^{-1}
\end{equation}

\begin{equation}
\label{eq:app_E}
    \bar{E} = \left( \frac{1-\nu_i^2}{E_i} + \frac{1-\nu_j^2}{E_j} \right) ^{-1}
\end{equation}

\begin{equation}
\label{eq:app_G}
    \bar{G} = \left( \frac{2(2 + \nu_i)(1 - \nu_i)}{E_i} + \frac{2(2 + \nu_j)(1 - \nu_j)}{E_j} \right) ^{-1}
\end{equation}

The force calculations associated with the Hooke model are given in Eq. \ref{eq:app_hooke}, where $v_c$ is the characteristic collision velocity. Additional details on the stiffness and damping parameters are available in \citet{zhang1996}. 

\begin{align}
\label{eq:app_hooke}
\begin{aligned}
    \vec{F}_n &= k_n \delta_n \hat{n} - g_n \vec{v}_n  \\ 
    \vec{F}_t &= \ -k_t \vec{\delta}_t - g_t \vec{v}_{t} \\
    k_n &= \frac{16}{15} \bar{E} \sqrt{R} \left( \frac{15 M v_c^2}{16 \bar{E} \sqrt{R}} \right)^{\frac{1}{5}}  \\
    k_t &= k_n \\
    g_n &= \sqrt{\frac{4 M k_n}{1 + \left( \pi / \ln{(e)} \right)^2}} \\
    g_t &= g_n \\ 
\end{aligned}
\end{align}

The force calculations associated with the Hertz model are given in Eq. \ref{eq:app_hertz}. Additional details on the stiffness and damping parameters are available in \citet{tsuji1992}.  

\begin{align}
\label{eq:app_hertz}
\begin{aligned}
    \vec{F}_n &= k_n \delta_n^{\frac{3}{2}} \hat{n} - g_n \delta_n^{\frac{1}{4}} \vec{v}_n \\ \vec{F}_t &= \ -k_t \delta_n^{\frac{1}{2}} \vec{\delta}_t - g_t \delta_n^{\frac{1}{4}} \vec{v}_{t} \\
    \noalign{\smallskip}
     k_n &= \frac{4}{3} \bar{E} \sqrt{R} \\
     k_t &= 8 \ \bar{G} \sqrt{R} \\
     \noalign{\smallskip}
     g_n &= \frac{-2 \ln(e)}{\sqrt{\ln^2(e) + \pi^2}} \sqrt{\frac{5}{6}} \sqrt{\frac{3}{2} M k_n} \\
     g_t &= \frac{-2 \ln(e)}{\sqrt{\ln^2(e) + \pi^2}} \sqrt{\frac{5}{6}} \sqrt{M k_t} \\
\end{aligned}
\end{align}

The force calculations associated with the \citet{flores2011} model are given in Eq. \ref{eq:app_flores}, where $c_n$ is the hysteresis damping factor and $v_{o}$ is the initial relative contact velocity between the spheres. The tangential force for the \citet{flores2011} model is the same as the tangential force for the Hertz model. 

\begin{align}
\label{eq:app_flores}
\begin{aligned}
    \vec{F}_n &= k_n \delta_n^{\frac{3}{2}} \hat{n} - c_n \delta_n^{\frac{3}{2}} \vec{v}_n \\ \vec{F}_t &= \ -k_t \delta_n^{\frac{1}{2}} \vec{\delta}_t - g_t \delta_n^{\frac{1}{4}} \vec{v}_{t} \\
    \noalign{\smallskip}
     k_n &= \frac{4}{3} \bar{E} \sqrt{R} \\
     k_t &= 8 \ \bar{G} \sqrt{R} \\
     \noalign{\smallskip}
     c_n &= \frac{8}{5}\ \frac{k_n\ (1 - e)}{e\ v_{o}}   \\
     g_t &= \frac{-2 \ln(e)}{\sqrt{\ln^2(e) + \pi^2}} \sqrt{\frac{5}{6}} \sqrt{M k_t} \\
\end{aligned}
\end{align}


\bsp	
\label{lastpage}
\end{document}